\documentclass[sigplan,natbib=false,balance=false]{acmart}
\newtoggle{draft}
\togglefalse{draft}

\newtoggle{extended}
\togglefalse{extended}

\usepackage[utf8]{inputenc}
\usepackage{cleveref}
\usepackage{xspace}

\usepackage{tolcolors}

\usepackage{float}
\usepackage{multirow}

\usepackage{xparse}

\iftoggle{draft}{\usepackage{parskip}}

\usepackage[natbib=true,maxcitenames=2,style=ACM-Reference-Format,sortcites=true]{biblatex}
\newcommand{\citenamed}[2][]{\citeauthor{#2}~\cite[#1]{#2}}
\newcommand{\Citenamed}[2][]{\Citeauthor{#2}~\cite[#1]{#2}}

\DeclareMathAlphabet\mathbfcal{OMS}{cmsy}{b}{n}

\usepackage{adjustbox}
\usepackage{framed}

\usepackage{array}
\newcolumntype{O}{>\!c<\!} 
\newcolumntype{T}{>$r<$}

\newcommand{\setOf}[1]{\{\, #1\, \}}

\let\mathscr\relax
\usepackage[scr]{rsfso}
\newcommand{\powerset}[1]{\mathscr{P}(#1)}

\newcommand{\alt}{\; | \;}

\usepackage{scopegraphs}
\DeclareMathAlphabet\mathbfcal{OMS}{cmsy}{b}{n}

\newcommand{\Scopes}{\ensuremath{\mathcal{S}}}
\newcommand{\scope}{\ensuremath{s}}
\newcommand{\scopes}{\ensuremath{S}}

\newcommand{\Labels}{\ensuremath{\mathcal{L}}}
\newcommand{\labels}{\ensuremath{L}}
\newcommand{\edgelbl}{\ensuremath{l}}
\newcommand{\EOP}{\ensuremath{\mathdollar}}
\newcommand{\PathLabels}{\ensuremath{\hat{\Labels}}}
\newcommand{\pathlbl}{\ensuremath{\hat{\edgelbl}}}
\newcommand{\pathlbls}{\ensuremath{\hat{L}}}

\newcommand{\Edges}{\ensuremath{\mathcal{E}}}
\newcommand{\edge}[3]{\ensuremath{#1\mkern-1mu\cdot\mkern-1mu#2\mkern-1mu\cdot\mkern-1mu#3}}
\newcommand{\edges}{\ensuremath{E}}

\newcommand{\Data}{\ensuremath{\mathcal{D}}}
\newcommand{\datum}{\ensuremath{d}}
\newcommand{\dataassoc}{\ensuremath{\rho}}

\newcommand{\SG}{\ensuremath{\mathcal{G}}}
\newcommand{\SGrep}[3]{\ensuremath{\langle #1, #2, #3 \rangle}}
\newcommand{\scopesOf}[1]{\ensuremath{\scopes_{#1}}}
\newcommand{\edgesOf}[1]{\ensuremath{\edges_{#1}}}

\newcommand{\assocOf}[2][\SG]{\ensuremath{\dataassoc_{#1}(#2)}}

\newcommand{\Paths}{\ensuremath{\mathcal{P}}}
\newcommand{\sgpath}{\ensuremath{p}}
\newcommand{\pathseg}[3]{\edge{#1}{#2}{#3}}

\newcommand{\AnswerSets}{\ensuremath{\mathcal{A}}}
\newcommand{\Answer}{\ensuremath{A}}

\newcommand{\source}[1]{\ensuremath{\mathsf{src}(#1)}}
\newcommand{\target}[1]{\ensuremath{\mathsf{tgt}(#1)}}

\newcommand{\re}{\ensuremath{R}}
\newcommand{\RE}{\ensuremath{\mathbfcal{R}}}
\newcommand{\derive}[1]{\ensuremath{\partial_#1}}
\newcommand{\rehead}[1]{\ensuremath{\mathscr{H}(#1)}}
\newcommand{\langOf}[1]{\ensuremath{\mathscr{L}(#1)}}

\newcommand{\reeps}{\ensuremath{\varepsilon}}

\newcommand{\reclos}[1]{\ensuremath{#1^\ast}}
\newcommand{\reopt}[1]{\ensuremath{#1?}}

\newcommand{\wfd}{\ensuremath{\mathbf{D}}}
\newcommand{\Wfd}{\ensuremath{\mathbfcal{D}}}

\newcommand{\lblOrd}{\ensuremath{<_{\hat{\edgelbl}}}}
\newcommand{\LblOrd}{\ensuremath{\mathcal{O}}}
\newcommand{\lblLE}{<}

\newcommand{\equivd}{\ensuremath{\approx_\datum}}
\newcommand{\Equivd}{\ensuremath{\mathbfcal{E}}}
\newcommand{\equivdOf}[2]{\ensuremath{#1 \equivd #2}}

\newcommand{\qrep}[6]{\ensuremath{\mathbf{query}\; #1, #2, #3, #4 \mathbf{in}\; #5 \mapsto #6}}
\newcommand{\cqrep}[5]{\ensuremath{\mathbf{query}\; #1, #2, #3 \mathbf{in}\; #4 \mapsto #5}}

\newcommand{\arrayOf}[1]{\overline{#1}}

\newcommand{\ResVars}{\ensuremath{\mathcal{X}}}
\newcommand{\resVar}{\ensuremath{x}}

\newcommand{\ResExps}{\ensuremath{\mathcal{E}}}
\newcommand{\resExp}{\ensuremath{E}}

\newcommand{\StateIds}{\ensuremath{\mathcal{N}}}
\newcommand{\stateId}{\ensuremath{n}}

\newcommand{\States}{\ensuremath{\mathcal{Y}}}
\newcommand{\State}{\ensuremath{Y}}

\newcommand{\StateMachines}{\ensuremath{\mathcal{M}}}
\newcommand{\StateMachine}{\ensuremath{M}}

\newcommand{\kw}[1]{\ensuremath{\mathbf{#1}}\xspace}
\newcommand{\kwResolve}{\kw{resolve}}
\newcommand{\kwSubenv}{\kw{subenv}}
\newcommand{\kwMerge}{\kw{merge}}
\newcommand{\kwShadow}{\kw{shadow}}

\newcommand{\kwState}{\kw{state}}
\newcommand{\kwStateMachine}{\kw{state\;machine}}

\newcommand{\kwElse}{\kw{else}}

\newcommand{\expResolve}{\ensuremath{\kwResolve}}
\newcommand{\expSubenv}[2]{\ensuremath{\kwSubenv \; #1 \; #2}}
\newcommand{\expMerge}[1]{\ensuremath{\kwMerge \; #1}}
\newcommand{\expShadow}[2]{\ensuremath{\kwShadow \; #1 \; #2}}
\newcommand{\expElse}[2]{\ensuremath{#1 \; \kwElse \; #2}}

\newcommand{\expState}[2]{\kwState\; #1}
\newcommand{\expStateMachine}[2]{\kwStateMachine\; #1}

\newcommand{\rulename}[1]{\textsc{#1}\xspace}
\newcommand{\ruleExpResolve}{\rulename{Exp-Resolve}}
\newcommand{\ruleExpSubenv}{\rulename{Exp-Subenv}}
\newcommand{\ruleExpMerge}{\rulename{Exp-Merge}}
\newcommand{\ruleExpShadow}{\rulename{Exp-Shadow}}
\newcommand{\ruleEvalState}{\rulename{Eval-State}}
\newcommand{\ruleResolveQuery}{\rulename{Op-Query-SM}}
\newcommand{\ruleExpElseL}{\rulename{Exp-Else-L}}
\newcommand{\ruleExpElseR}{\rulename{Exp-Else-R}}

\newcommand{\envstore}{\Sigma}

\NewDocumentCommand{\expsem}{ O{\envstore} O{\StateMachine} O{\SG} O{\sgpath} O{\wfd} O{\equivd} m m}{#1, #2, #3, #4, #5, #6\; |- #7 => #8}
\NewDocumentCommand{\statesem}{ O{\StateMachine} O{\SG} O{\sgpath} O{\wfd} O{\equivd} m m}{#1, #2, #3, #4, #5\; |- #6 => #7}
\newcommand{\envvalue}[2][\envstore]{#1(#2)}
\newcommand{\sminit}[1]{\mathsf{init}(#1)}

\usepackage{semantic}
\usepackage{mathtools}

\makeatletter
\newcommand\@eatpar{\@ifnextchar\par{\expandafter\@eatpar\@gobble}\relax}
\newcommand{\figuresection}[2][]{%
  \par%
  {\sffamily\bfseries #2}\hfill{#1}%
  \smallskip%
  \@eatpar}
\makeatother

\usepackage[noend, linesnumbered, commentsnumbered]{algorithm2e}
\usepackage{boxedminipage}

\makeatletter
\newcommand{\removelatexerror}{\let\@latex@error\@gobble}
\makeatother

\newenvironment{boxedalgorithm}[1][\linewidth]{
    \removelatexerror
    \begin{boxedminipage}[t]{#1}
    \begin{algorithm}[H]
}{
    \end{algorithm}
    \end{boxedminipage}
}

\DontPrintSemicolon
\SetKwProg{Fn}{fun}{}{end fun}
\SetKwProg{Block}{}{}{}
\newcommand{\withRT}[1]{$: #1$}
\newcommand{\Ternary}[3]{\leIf{#1}{#2}{#3}\par}
\SetKwInOut{Input}{input}
\SetKwInOut{Output}{output}

\def\localfundefskip{0.25em}
\SetKw{LocalFnTitle}{fun}
\newcommand{\LocalFn}[2]{\LocalFnTitle #1$\triangleq$ #2\vspace{\localfundefskip}}
\newcommand{\EndLocalFnDef}{\;}

\SetKwFunction{ResolveOrdered}{TraverseOrdered}%
\SetKwFunction{Traverse}{Traverse}%

\newcommand{\hatL}{\mkern-3.5mu\hat{\mkern4mu\mathsf{L}}}

\newcommand{\withRTNR}[1]{}
\SetKwFunction{Resolve}{Resolve}
\SetKwFunction{ResolveEOP}{Resolve-$\mathsf{\EOP}$}
\SetKwFunction{Resolvel}{Resolve-$\mathsf{\edgelbl}$}
\SetKwFunction{Resolvell}{Resolve-$\hat{\mathsf{l}}$}
\SetKwFunction{ResolveLl}{Resolve-$\hat{\mathsf{l}}\hatL$}
\SetKwFunction{ResolveL}{Resolve-$\hatL$}
\SetKwFunction{ResolveAll}{Resolve-All}
\SetKwFunction{Shadow}{Shadow}

\newcommand{\maxOf}[2][\lblOrd]{\ensuremath{\mathsf{max}_{#1}(#2)}}
\newcommand{\smallerOf}[3][\lblOrd]{\ensuremath{\mathsf{smaller}_{#1}(#2, #3)}}

\colorlet{query-1}{colorblind-muted-7}
\colorlet{query-2}{colorblind-muted-5}
\colorlet{query-3}{colorblind-muted-3}
\colorlet{query-4}{colorblind-muted-8}

\newcommand{\coloredref}[2]{\texttt{\textcolor{#1}{#2}}}

\usepackage{listings}

\lstdefinestyle{defaultstyle}{
  basicstyle=\ttfamily\small,
  showstringspaces=false,
  commentstyle=\color{green!50!black},
  keywordstyle=\bfseries
}

\lstdefinelanguage{PCF}{
  keywords={let,fun,in,true,false},
  moredelim={**[is][\color{query-1}]{@1}{@}},
  moredelim={**[is][\color{query-2}]{@2}{@}},
  moredelim={**[is][\color{query-3}]{@3}{@}},
  moredelim={**[is][\color{query-4}]{@4}{@}},
}

\lstdefinelanguage{LM}{
  keywords={module,import,def,fun,fix,let,letrec,letpar,in},
  moredelim={**[is][\color{query-1}]{@1}{@}},
  moredelim={**[is][\color{query-2}]{@2}{@}},
  moredelim={**[is][\color{query-3}]{@3}{@}},
  moredelim={**[is][\color{query-4}]{@4}{@}},
}

\lstdefinelanguage{SM}{
  keywords={state,machine,resolve,subenv,merge,shadow},
  moredelim={**[is][\sffamily]{@lbl@}{@}},
}

\newcommand{\lbldef}[1]{\ensuremath{\mathsf{#1}}\xspace}

\newcommand{\lblMOD}{\lbldef{MOD}}
\newcommand{\lblVAR}{\lbldef{VAR}}

\newcommand{\lblLEX}{\lbldef{P}}
\newcommand{\lblIMPORT}{\lbldef{I}}

\newcommand{\lblL}{\lbldef{L}}

\newcommand{\with}{\mapsto}
\newcommand{\id}[1]{\mathtt{#1}}

\newcommand{\tyFUN}[2]{\id{\boldmath #1 \to #2}}

\newcommand{\tyNAT}{\mathbb{N}}

\usepackage{calc}
\newcommand{\cidx}[1]{_{\makebox[\widthof{\small\texttt{x}}]{\scriptsize\ensuremath{#1}}}}

\newenvironment{compilationexample}[3][0.33\linewidth]{
  \begin{adjustbox}{minipage={#1}, padding=0ex 0ex, frame, max width = #1}%
    \begin{adjustbox}{minipage={0.5\linewidth}, padding=1ex 1ex, max width = 0.5\linewidth}%
      \centering
      $\re{:}\; #2$\vphantom{\lblOrd\EOP}
    \end{adjustbox}%
    \vrule%
    \begin{adjustbox}{minipage={0.5\linewidth}, padding=1ex 1ex, max width = 0.5\linewidth}%
      \centering
      $\lblOrd:\, #3$
    \end{adjustbox}
    \hrule
    \begin{adjustbox}{minipage={\linewidth}, padding=1ex 0ex, max width = \linewidth}%
}{%
    \end{adjustbox}%
  \end{adjustbox}%
}

\newcommand{\sid}{\mathsf{id}}
\newcommand{\sidT}{\Labels \rightharpoonup \StateIds}
\newcommand{\sidOf}[1]{\sid(#1)}

\newcommand{\semName}[1]{\mathsf{spec}_{#1}}
\newcommand{\sem}[2]{#1(#2)}

\newcommand{\semSM}{\semName{\mathsf{\StateMachine}}}
\newcommand{\semSMOf}[2]{\sem{\semSM}{#1, #2}}

\newcommand{\semY}{\semName{\mathsf{\State}}}
\newcommand{\semYOf}[3]{\sem{\semY}{#1, #2, #3}}

\newcommand{\semL}{\semName{\mathsf{L}}}
\newcommand{\semLOf}[3]{\sem{\semL}{#1, #2, #3}}

\newcommand{\semLl}{\semName{\mathsf{lL}}}
\newcommand{\semLlOf}[4]{\sem{\semLl}{#1, #2, #3, #4}}

\newcommand{\seml}{\semName{\mathsf{l}}}
\newcommand{\semlOf}[2]{\sem{\seml}{#1, #2}}

\newcommand{\stmConcat}{\mathrel{\oplus}}
\newcommand{\StmConcat}{\mathlarger{\oplus}\,}

\newcommand{\cquote}[1]{#1}
\newcommand{\cunquote}[1]{#1}

\newcommand{\where}{\;\mathit{where}}

\newcommand{\genStates}{\mathsf{gen\_states}}
\newcommand{\genStatesOf}[1]{\genStates(#1)}

\newcommand{\fv}{\mathsf{fresh\_id}}
\newcommand{\freshVar}{\fv()}

\newcommand{\unzip}{\mathsf{unzip}}
\newcommand{\unzipOf}[1]{\unzip(#1)}

\newcommand{\tuple}[1]{\langle #1 \rangle}

\title{Specializing Scope Graph Resolution Queries}

\author{Aron Zwaan}

\orcid{0000-0002-1818-4245}
\affiliation{
  \department{Software Technology}
  \institution{Delft University of Technology}
  \city{Delft}
  \country{Netherlands}
}
\email{a.s.zwaan@tudelft.nl}


\setcopyright{rightsretained}
\acmPrice{15.00}
\acmDOI{10.1145/3567512.3567523}
\acmYear{2022}
\copyrightyear{2022}
\acmSubmissionID{splashws22slemain-p36-p}
\acmISBN{978-1-4503-9919-7/22/12}
\acmConference[SLE '22]{Proceedings of the 15th ACM SIGPLAN International Conference on Software Language Engineering}{December 06--07, 2022}{Auckland, New Zealand}
\acmBooktitle{Proceedings of the 15th ACM SIGPLAN International Conference on Software Language Engineering (SLE '22), December 06--07, 2022, Auckland, New Zealand}


\begin{document}

\begin{abstract}
To warrant programmer productivity, type checker results should be correct and available quickly.
Correctness can be provided when a type checker implementation corresponds to a declarative type system specification.
Statix is a type system specification language which achieves this by automatically deriving type checker implementations from declarative typing rules.
A key feature of Statix is that it uses scope graphs for declarative specification of name resolution.
However, compared to hand-written type checkers, type checkers derived from Statix specifications have sub-optimal run time performance.

In this paper, we identify and resolve a performance bottleneck in the Statix solver, namely part of the name resolution algorithm, using partial evaluation.
To this end, we introduce a tailored procedural intermediate query resolution language, and provide a specializer that translates declarative queries to this language.

Evaluating this specializer by comparing type checking run time performance on three benchmarks (Apache Commons CSV, IO, and Lang3), shows that our specializer improves query resolution time up to 7.7x, which reduces the total type checking run time by 38 -- 48\%.
\end{abstract}

\maketitle

\section{Introduction}
\label{sec:introduction}

Developers, whether they use a general-purpose or a domain-specific language (DSL), use static name and type analysis to understand and evolve their code.
However, implementing a type checker takes significant time and effort.
In particular, implementing name binding correctly is challenging, as it requires careful staging of program traversals~\cite{RouvoetAPKV20}.
Therefore, type checker frameworks that abstract over name resolution scheduling, such as~\citenamed{PacakES20} (based on Datalog), \citenamed{WykBGK10}, \citenamed{HedinM03} (using attribute grammars), and \Citenamed{AntwerpenPRV18} (using constraint programming and scope graphs) have been developed.
These frameworks ensure executable type checkers can be developed with significantly reduced effort.

Interpreting such declarative specifications often requires intricate logic.
Generally, the more a language abstracts from implementation details, the more complicated an interpreter or compiler will be.
However, this comes with the risk of introducing significant run time overhead, resulting in suboptimal performance compared to low-level approaches.

In this paper, we improve the performance of type checkers based on \emph{scope graphs}~\cite{NeronTVW15, AntwerpenPRV18}.
Scope graphs are an established approach to modeling name-binding structure.
In this model, the scoping structure and declarations of a program are represented in a graph.
References can be resolved using a versatile graph query mechanism.
Scope graphs are embedded in the Statix DSL for type system specification~\cite{AntwerpenPRV18, RouvoetAPKV20}.
This DSL allows high-level specification of type systems using declarative inference rules.
It has a well-defined declarative semantics, which allows reasoning over type-systems while abstracting over operational details.
The Statix solver interprets specifications as constraint programs, which yields executable type checkers that are sound with respect to the declarative semantics.
Case studies using Statix have shown that scope graphs are expressive enough to support type systems with non-lexical bindings (e.g., imports and inheritance), structural types, and parametric polymorphism~\cite{NeronTVW15, AntwerpenPRV18}.
In addition, they allow language-parametric definition of editor services, such as semantic code completion~\cite{PelsmaekerAPV22}, renaming~\cite{Misteli20,Misteli21} and inlining~\cite{Gugten22}.
The expressiveness, declarativity, and additional services makes it especially suitable for DSLs and rapid language prototyping.

However, 
type checkers derived from Statix specifications are rather slow.
For example, type checking the Apache Commons IO library takes 14.7 seconds with the concurrent solver using 8 cores and even 73.4 seconds on a single core~\cite{AntwerpenV21-artifact}.
On the same machine, a full compilation using \texttt{javac} takes roughly 3 seconds on 8 cores, and 5 seconds on a single core.

In this paper, 
we apply partial evaluation to resolve a newly identified performance bottleneck in Statix' scope graph query resolution algorithm.
Our evaluation shows that the approach ensures query resolution is up to 7.7x faster than traditional query resolution on average.
This improves the performance of Statix-based type checkers by 38 -- 48\% on Java projects, which is a significant step forward to applying generated type checkers on larger codebases.

In summary, our contributions are as follows:
\begin{itemize}


  \item We explain the scope graph query resolution algorithm, and identify a performance bottleneck in the algorithm (\cref{sec:query-resolution-in-scope-graphs}).

  \item We introduce an intermediate language that makes scope graph traversal order and partial query result combination explicit (\cref{sec:query-resolution-language}).

  \item We present a specializer from traditional scope graph queries to our new intermediate language (\cref{sec:compiling-declarative-queries}).

  \item We evaluate the correctness and performance of our approach (\cref{sec:evaluation}).
        We show that specializing queries makes scope graph query resolution up to 7.7x faster.

\end{itemize}

\section{Partial Evaluation for DSL Interpreters}
\label{sec:partial-evaluation}

In this section, we provide a brief introduction to partial evaluation (popularized by \citenamed{Futamura82}), and explain why we think it is especially beneficial for declarative languages such as Statix.
From the perspective of partial evaluation, a program can be seen as a function from inputs to an output $O$.
Some of these inputs may be known statically ($S$), while some of them may vary per invocation ($D$).
Then, the signature of a program looks as follows:
\[
  \mathsf{prog} : S \times D \to O
\]
A \emph{specializer} takes such a program and its static input, and returns a \emph{residual program} $\mathsf{prog}_{S} : D \to O$.
When generating $\mathsf{prog}_{S}$, it performs the part of the computation that does not depend on $D$, making $\mathsf{prog}_{S}$ generally faster than $\mathsf{prog}$.

\subsection{Partial Evaluation for Interpreters}

This pattern can easily be applied to programming languages.
In that case, $\mathsf{prog}$ is an interpreter that evaluates the semantics of a program (its static input $S$) with respect to some arguments $D$.
The residual program is essentially a compiled version of $S$.
This is called the \emph{first Futamura projection}.

Specialization is generally only beneficial when a program is executed multiple times.
However, Futamura argues that specializing an interpreter to a program may already be beneficial when executing the program once, as programs may repeatedly evaluate a particular piece of code.
Specializing repeatedly executed program fragments removes the interpretation overhead, which might outweigh the run time costs of compilation.
This effect becomes stronger when the computational complexity of interpreting particular language constructs is high.
That is, the more overhead an interpreter introduces, the more beneficial specialization will be.

Declarative languages are languages in which a user specifies intent rather than procedure.
The logic to compute a result that corresponds to the intent is then implemented in the interpreter (or compiler) of the language.
Thus, a declarative language moves part of the complexity of a problem from the program or programmer to its interpreter.

Having an interpreter with intricate logic means that declarative languages are susceptible to introducing relatively more run time overhead compared to non-declarative languages.
Interpreters of declarative languages might have to execute non-trivial algorithms in order to evaluate a program.
For that reason, partial evaluation might be relatively beneficial for declarative languages.

\subsection{Application to Statix}

Applying partial evaluation to Statix introduces a few problems.
In fact, it is as complex as finding a compiler from a constraint language with an embedded scope graph logic to an imperative language, such as Java.
Such a compiler should ensure that all internal scheduling of rule selection and query resolution is handled correctly.
We regard this as an open research challenge that is too complicated to solve in one step.
Instead, in this paper, we specialize only a computationally complex part of the interpreter, namely the query resolution algorithm, to a specification.
This yields a specification in which the query resolution constraints are partially evaluated, but other constraints are not.
This specification can then be interpreted by a constraint solver without using the query resolution algorithm, ensuring faster execution.

To characterize this approach, regard a Statix specification $C_Q \uplus C_O$ as a collection of query constraints $C_Q$ and other constraints $C_O$.
The $\uplus$ symbol indicates that these groups are mutually embedded in actual specifications.
Our approach is then summarized in the following functions:
\begin{align*}
  \mathsf{specialize}  &: C_Q \uplus C_O \to C_Q^\ast \uplus C_O \\
  \mathsf{solve} &: C_Q^\ast \uplus C_O \times P \to T
\end{align*}
Here $\mathsf{specialize}$ specializes the query resolution algorithm with respect to particular query constraints, yielding specialized queries ($C_Q^\ast$).
These constraints are embedded in the original AST, yielding a partially specialized specification $C_Q^\ast \uplus C_O$.
When type-checking a concrete object program $P$, this specialized specification is interpreted by an adapted solver $\mathsf{solve}$, returning a typing $T$.

These specialized queries $C_Q^\ast$ cannot be represented in either Statix itself or Java (the language in which the solver is written).
Thus, we introduce a tailored intermediate language to express those.
For this language, we provide an interpreter, which the Statix solver uses instead of the name resolution algorithm.

\section{Resolving Queries in Scope Graphs}
\label{sec:query-resolution-in-scope-graphs}

In this section, we introduce scope graphs and query resolution.
\Cref{subsec:query-resolution-example} introduces scope graphs and three parameters of scope graph queries using two examples.
After that, \cref{subsec:algorithm-outline} explains how Statix interprets these query parameters.
Then, we explain that repeated querying of some of these parameters makes query resolution slow, which motivates the effort to optimize it (\cref{subsec:resolution-performance}).
Finally, \cref{subsec:resolution-algorithm} provides the full resolution algorithm, which is required to understand the remainder of the paper.

\subsection{Query Resolution by Example}
\label{subsec:query-resolution-example}

We consider two examples of queries in scope graphs.
These examples motivate the declarative query language of Statix, as well as explaining the resolution algorithm that interprets such queries.
Each example discusses the scope graph of a program, and the resolution of a particular query in that scope graph.

\begin{figure}
  \begin{adjustbox}{minipage={\linewidth}, padding=2pt, frame, max width = \linewidth}

    \centering
    \begin{tikzpicture}[scopegraph, node distance = 1.5em and 2em]
      \node[scope] (s) {$s_{l}$};

      \node[scope, right = of s] (s-x) {$s_x \with \id{x} : \tyNAT$};
      \draw (s) edge[lbl=\lblVAR] (s-x);

      \node[scope, below right = of s] (s-lmb) {$s_{\lambda}$};
      \draw (s-lmb) edge[lbl=\lblLEX] (s);

      \node[scope, color = white, below = of s-lmb] (inv) {};

      \node[scope, below left = of s-lmb] (s-l) {$s_{l'}$};
      \draw (s-l) edge[lbl=\lblLEX] (s);

      \node[scope, left = of s-l] (s-f) {$s_f \with \id{f} : \tyFUN{\tyNAT}{\tyNAT}$};
      \draw (s-l) edge[lbl=\lblVAR] (s-f);

      \node[scope, below = 3em of s-lmb.west, anchor = west] (s-y) {$s_y \with \id{y} : \tyNAT$};
      \draw (s-lmb) edge[lbl=\lblVAR] ([xshift=-1.4em]s-y.north);

      \node[below left = -1em and 1em of s] (code) {
        \begin{lstlisting}[language=PCF]
let x = 6 in
let f = fun y. @1x@ * y
 in f 7
        \end{lstlisting}
      };

      \begin{scope}[color = query-1, dashed]
        \node[scope, right = of s-lmb] (q-x) {$\id{x}$};

        \draw (q-x) edge[] (s-lmb);

        \draw (s-lmb) edge[bend left, lbl=1]      (s-y);
        \draw (s-lmb) edge[bend left = 20, lbl=2] (s);
        \draw (s)     edge[bend left, lbl=3]      (s-x);
      \end{scope}

    \end{tikzpicture}
  \end{adjustbox}

  \caption{PCF program with Scope Graph and Query}
  \label{fig:example-pcf}

\end{figure}
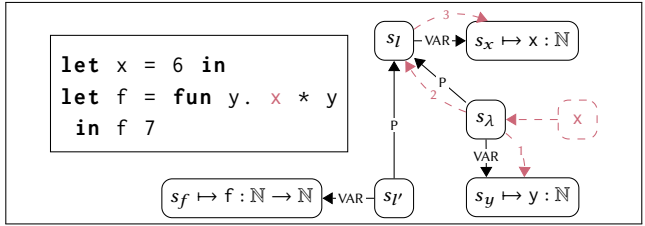


\paragraph{Example 1}
In \cref{fig:example-pcf}, a small PCF program, containing two let-bindings, a function definition and its application, is shown.
Next to the program, the scope graph is depicted.
In a scope graph, each `scope' is modeled by a node.
In this case, $s_l$ and $s_{l'}$ represent the bodies of the let expressions, while $s_\lambda$ models the body of the function.
The \lblLEX-labeled edges, such as the edge from $s_{l'}$ to $s_l$, ensure declarations of outer scopes will be visible in inner scopes.
Nodes $s_x$, $s_y$, and $s_f$ model the declarations of the \texttt{x}, \texttt{y}, and \texttt{f} variables, respectively.
Therefore, these scopes map to a datum (e.g. $\id{x} : \tyNAT$ for $s_x$) that indicates the name and type of the declaration.

References are modeled using \emph{queries} in scope graphs.
In the code, a reference \coloredref{query-1}{x} is highlighted.
This reference corresponds to the dashed box in the scope graph.
The box points to $s_\lambda$, because \coloredref{query-1}{x} occurs in the body of the function expression.
Eventually, the query resolves to $s_x$ via $s_l$, which is indeed the declaration of \texttt{x} in the outer let.

For this paper, we are particularly interested in \emph{how} this query result was computed.
This is indicated by the numbered, dashed edges.
When starting the query in $s_\lambda$, the algorithm first traverses the \lblVAR edge to $s_y$.
Then it checks whether $s_y$ is a valid declaration for the reference.
Since this is not the case, the algorithm continues by traversing the \lblLEX edge to $s_l$ (step 2).
From there, the \lblVAR edge to $s_x$ is traversed.
The algorithm finds that $s_x$ is a valid declaration, and returns that as the environment to which the query evaluates.

In Statix, one would not write a resolution procedure as shown above directly, as Statix is meant to be a declarative specification language.
Instead, Statix interprets a high-level description of valid query answers using a \emph{generic} algorithm, yielding the behavior as shown above.
So how can we describe the query resolution procedure in a high level fashion?

The given example already shows two of the three \emph{query parameters} that determine how a query resolves to an environment.
First, the query resolution algorithm decided that $s_y$ should not be in the environment, while $s_x$ should.
That is expressed using a \emph{data well-formedness condition} $\wfd$, which is a unary predicate over datums.
A possible declaration is only included in the environment when its datum matches $\wfd$.
In this case, the predicate only accepts declarations with name $\id{x}$.
Second, the algorithm decided to traverse \lblVAR edges before \lblLEX edges.
This corresponds to the intuition that local declaration are prefered over (i.e., shadow) more distant declarations.
In Statix, this is modeled using a strict partial order over labels (refered to as \emph{label order}).
For this example, the label order $\lblVAR \lblLE \lblLEX$ indicates that \lblVAR edges should be traversed first.



\paragraph{Example 2}

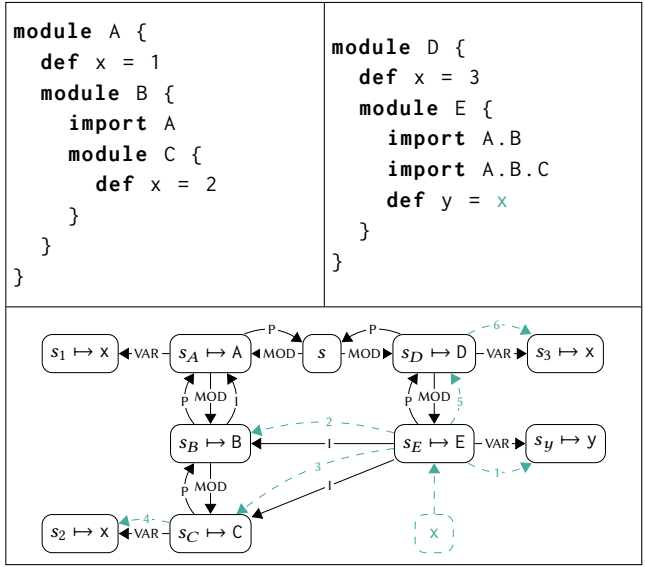
\begin{figure}
  \begin{adjustbox}{minipage={\linewidth}, padding=0pt, frame, max width = \linewidth}
    \begin{adjustbox}{minipage={0.5\linewidth}, padding=2pt, max width = 0.5\linewidth}
      \begin{lstlisting}[language=LM]
module A {
  def x = 1
  module B {
    import A
    module C {
      def x = 2
    }
  }
}
      \end{lstlisting}
    \end{adjustbox}%
    \vrule%
    \begin{adjustbox}{minipage={0.5\linewidth}, padding=2pt, max width = 0.5\linewidth}
      \begin{lstlisting}[language=LM]
module D {
  def x = 3
  module E {
    import A.B
    import A.B.C
    def y = @3x@
  }
}
      \end{lstlisting}
    \end{adjustbox}%

    \hrule
    \begin{adjustbox}{minipage={\linewidth}, padding=1ex, max width = \linewidth}
      \centering
      \begin{tikzpicture}[scopegraph, node distance = 2em and 2em]

        \node[scope] (s) {$s$};

        \node[scope, left = of s]                         (s-A) {$s_A \with \id{A}$};
        \draw (s)   edge[lbl=\lblMOD]                     (s-A);
        \draw (s-A) edge[lbl=\lblLEX, bend left]          (s);

        \node[scope, left = of s-A]                       (s-1) {$s_1 \with \id{x}$};
        \draw (s-A) edge[lbl=\lblVAR]                     (s-1);

        \node[scope, below = of s-A]                      (s-B) {$s_B \with \id{B}$};
        \draw (s-A) edge[lbl=\lblMOD]  (s-B);
        \draw (s-B) edge[lbl=\lblLEX, bend left = 40]     (s-A);
        \draw (s-B) edge[lbl=\lblIMPORT, bend right = 40] (s-A);

        \node[scope, below = of s-B]                      (s-C) {$s_C \with \id{C}$};
        \draw (s-B) edge[lbl=\lblMOD]                     (s-C);
        \draw (s-C) edge[lbl=\lblLEX, bend left = 40]     (s-B);

        \node[scope, left = of s-C]                       (s-2) {$s_2 \with \id{x}$};
        \draw (s-C) edge[lbl=\lblVAR]                     (s-2);

        \node[scope, right = of s]                        (s-D) {$s_D \with \id{D}$};
        \draw (s)   edge[lbl=\lblMOD]                     (s-D);
        \draw (s-D) edge[lbl=\lblLEX, bend right]         (s);

        \node[scope, right = of s-D]                      (s-3) {$s_3 \with \id{x}$};
        \draw (s-D) edge[lbl=\lblVAR]                     (s-3);

        \node[scope, below = of s-D]                      (s-E) {$s_E \with \id{E}$};
        \draw (s-D) edge[lbl=\lblMOD]  (s-E);
        \draw (s-E) edge[lbl=\lblLEX, bend left = 40]     (s-D);
        \draw (s-E) edge[lbl=\lblIMPORT]                  (s-B);
        \draw (s-E) edge[lbl=\lblIMPORT]                  (s-C);

        \node[scope, right = of s-E]                      (s-y) {$s_y \with \id{y}$};
        \draw (s-E) edge[lbl=\lblVAR]                     (s-y);

        \begin{scope}[dashed, color = query-3]
          \node[scope, below = of s-E]                    (q)   {$\id{x}$};

          \draw (q)   edge[]                              (s-E);
          \draw (s-E) edge[lbl=1, bend right]             (s-y);
          \draw (s-E) edge[lbl=2, bend right = 15]        (s-B);
          \draw (s-E) edge[lbl=3, bend right = 15]        (s-C);
          \draw (s-C) edge[lbl=4, bend right = 15]        (s-2);
          \draw (s-E) edge[lbl=5, bend right = 40]        (s-D);
          \draw (s-D) edge[lbl=6, bend left]              (s-3);
        \end{scope}

      \end{tikzpicture}
    \end{adjustbox}
  \end{adjustbox}%

  \caption{LM Example. Variable types are omitted for brevity.}
  \label{fig:example-modules}

\end{figure}

\Cref{fig:example-modules} shows an example program in Language with Modules (LM, introduced by \citenamed{NeronTVW15}).
In this scope graph, scope $s$ represent the \emph{global scope}.
Scopes $s_A$, $s_B$, $s_C$, $s_D$ and $s_E$ have a double role: they model the declaration of a module as well as its body.
Therefore, they have incoming \lblMOD edges and a datum as well as inner declarations.
All modules have a \lblLEX edge back to their enclosing context.
Finally, the imports are modeled using \lblIMPORT edges.

When resolving \coloredref{query-3}{x} in $s_E$, the resolution algorithm first traverses the \lblVAR edge to $s_y$ (step 1).
Because that declaration does not match, it continues traversing the \lblIMPORT edge to $s_B$ (step 2).
That scope does not contain valid declarations but it has \lblLEX and \lblIMPORT edges to $s_A$, and a \lblMOD edge to $s_C$.
However, reference resolution should not traverse those.
Clearly, traversing the \lblLEX edge would be incorrect.
After all, module $\id{E}$ only imports $\id{A.B}$, which should not bring declarations from $\id{A}$ in scope.
Whether traversing the \lblIMPORT edge is valid depends on the language.
Languages with transitive imports would allow traversing multiple subsequent \lblIMPORT edges, while language with non-transitive imports would not.
As LM has non-transitive imports, we do not traverse the \lblIMPORT edge to $s_A$.
Similarly, the \lblMOD edge to $s_C$ should not be traversed, as modules should not be imported implicitly.
Instead, the query resolves to declaration $s_2$ in $s_C$ via the import edge from $s_E$ (step 3 and 4).
Now, we still need to consider whether it is necessary to traverse the \lblLEX edge to $s_D$.
When imports shadow the surrounding scope, that is not required, as $s_2$ would shadow any result from $s_D$.
However, for the sake of the example, we assume that imports and the enclosing scope have equal priority.
Thus, $s_3$ is resolved as well (step 5 and 6), which means that reference \coloredref{query-3}{x} is ambiguous.

This example shows that there can be additional constraints on paths.
In Statix, these \emph{path-wellformedness conditions} are expressed as a \emph{regular expression} (RE) on path labels.
The query resolution algorithm only traverses paths that are valid with respect to the given RE.
In particular, the regular expression that describes this query is $\reclos{\lblLEX}\reopt{\lblIMPORT}\lblVAR$.
This RE allows looking in enclosing scopes ($\reclos{\lblLEX}$), possibly traversing a single import edge ($\reopt{\lblIMPORT}$), while ensuring that only variables are resolved ($\lblVAR$).

Second, we argued that the reference is ambiguous when imports and lexical enclosing scopes have equal priority.
This can be modeled by having neither $\lblIMPORT \lblLE \lblLEX$ nor $\lblLEX \lblLE \lblIMPORT$ in the label order.
Then, the label order does not provide any priority for a particular environment, and the resolution algorithm will return them both.

\paragraph{Summary}

These examples show how scope graphs can be used to model the name binding structure of PCF and LM programs.
Resolution of references is done using declarative scope graph queries.
Valid results of a query are described by three parameters.
First, the path well-formedness condition, expressed as a regular expression over labels, describes valid paths.
Second, the data well-formedness condition selects which declarations are valid.
Finally, the label order condition describes which declarations shadow each other.
A query resolution algorithm, integrated in the Statix solver, interprets such parameters to compute the result of a query.

\subsection{Query Resolution: Algorithm Outline}
\label{subsec:algorithm-outline}

When executing a Statix specification, a resolution algorithm resolves queries.
This algorithm performs an advanced depth-first search, using the aforementioned query parameters to find correct results.
In this section, we explain how each of the query parameters is used by the algorithm.

\paragraph{Path well-formedness}
First, consider the use of the path well-formedness RE $\re$.
The fact that valid paths need to adhere to $\re$ implies that query resolution does not have to traverse the whole graph, but only edges with labels that do not violate $\re$.
The labels $\edgelbl$ that require traversal are precisely those for which the Brzozowski derivative $\derive{\edgelbl} \re$~\cite{Brzozowski64} is not equal to the empty language $\varnothing$.
We call this set the \emph{head set} of a regular expression (written as $\rehead{\re}$), defined as:
\[ \rehead{\re} \triangleq \setOf{ \edgelbl \in \Labels \alt \derive{\edgelbl} \re \neq \varnothing} \]
Moreover, after traversing an edge with label $\edgelbl$, the resolution algorithm uses $\derive{\edgelbl} \re$ for further exploration from the target node.
This retains the invariant that full resolution paths will adhere to the initial regular expression.

To illustrate this, recall the example in \cref{fig:example-modules}.
The query in this example had $\re_2 = \reclos{\lblLEX}\reopt{\lblIMPORT}\lblVAR$ as initial RE.
Because this regular expression has derivatives with respect to \lblVAR, \lblIMPORT and \lblLEX, all edges were traversed (step 1, 2, 3 and 5).
However, from $s_B$, the \lblIMPORT and \lblLEX edges were not traversed because $\derive{\lblIMPORT} \re_2 = \lblVAR$ was used as RE after traversing the \lblIMPORT edge (step 2).

\paragraph{Label Order}
The label order $\lblOrd$ is used to model shadowing.
This is implemented in the algorithm as follows.
The algorithms traverses edges in topological ascending order of their labels.
Results obtained from a particular edge are then only included when they are not shadowed by results obtained from an edge with a smaller label.

As an example, consider the label order used in the LM example: $\lblVAR \lblLE \lblLEX, \lblVAR \lblLE \lblIMPORT$.
This order indicates that results over \lblVAR labels shadow results from \lblLEX and \lblIMPORT, but the latter two do not shadow each other.
Thus, the algorithm first traverses \lblVAR edges.
The result of that traversal is then used to shadow results obtained from \lblLEX and \lblIMPORT edges.
That is, when the \lblVAR traversal returned a non-empty environment, declarations reached via \lblLEX and \lblIMPORT edges are ignored.
However, as the label order is not necessarily total, some variability in the order of traversing mutually incomparable labels is possible.
In the above example, it does not matter whether \lblLEX edges are traversed before \lblIMPORT edges, or the other way around, but they should not \emph{shadow} each other.

Internally, the algorithm determines the label traversal order using two helper functions:
\begin{align*}
  \maxOf{\pathlbls} &\triangleq \setOf{ \pathlbl \in \pathlbls \alt \not\exists \pathlbl' \in \pathlbls.\; \pathlbl \lblOrd \pathlbl' }\\
  \smallerOf{\pathlbls}{\pathlbl} &\triangleq \setOf{ \pathlbl' \in \pathlbls \alt \pathlbl' \lblOrd \pathlbl }
\end{align*}
Here, $\mathsf{max}$ computes the labels that are last in the label order, which are hence traversed \emph{last}.
For each label $\pathlbl$ in the $\mathsf{max}$ set, its $\mathsf{smaller}$ set is computed.
This set contains precisely those labels that shadow the $\pathlbl$ label.
Using these functions, the actual traversal order and shadowing of a set of labels $\pathlbls$ is determined as shown in \cref{fig:edge-traversal-order}.
The \ResolveOrdered function receives a set of labels it must traverse in the correct order.
If the set is empty, an empty environment is returned (line 2), as no edges need to be traversed.
Otherwise, the algorithm initializes an empty environment $\Answer$, which eventually will contain all declarations reachable over labels in $\pathlbls$.
Then, it iterates over all labels $\pathlbl$ in the $\mathsf{max}$ set of $\pathlbls$.
For each $\mathsf{max}$ label $\pathlbl$, all declarations that may possibly shadow it are computed by recursively applying \ResolveOrdered on its smaller set.
This ensures labels in the $\mathsf{smaller}$ set are traversed before $\pathlbl$.
After that, all $\pathlbl$-labeled edges are traversed.
The result of the latter operation is shadowed against the result of the former, and the result is added to $\Answer$.

\begin{figure}
\begin{boxedalgorithm}
  \Fn{\ResolveOrdered{$\pathlbls$}}{
    \lIf{$\pathlbls = \emptyset$}{\Return $\emptyset$}
    $\Answer$ := $\emptyset$\;
    \ForEach{$\pathlbl \in \maxOf{\pathlbls}$}{
      $\Answer_L$ := \ResolveOrdered{$\smallerOf{\pathlbls}{\pathlbl}$}\;
      $\Answer_l$ := \Resolvell{$\pathlbl$}\;
      $\Answer$ += \Shadow{$\Answer_L$, $\Answer_l$}\;
    }
    \Return $\Answer$\;
  }
\end{boxedalgorithm}
\caption{Edge Traversal Order.}
\label{fig:edge-traversal-order}
\end{figure}

An astute reader might ask why we do not simply return $\Answer_L$, and only compute $\Answer_l$ when the former is empty.
This is caused by an additional query parameter we do not discuss here.
We return to this in \cref{subsec:resolution-algorithm,subsec:skipping-shadowed-environments}.

\begin{figure}
\begin{boxedalgorithm}
  \ResolveOrdered{$\setOf{\lblVAR, \lblIMPORT, \lblLEX}$}\;
  \Indp
    $A$ := $\emptyset$\;
    $\maxOf{\setOf{\lblVAR, \lblIMPORT, \lblLEX}} = \setOf{\lblIMPORT, \lblLEX}$\;
    \Indp
      $\smallerOf{\setOf{\lblVAR, \lblIMPORT, \lblLEX}}{\lblIMPORT} = \setOf{\lblVAR}$\;
      \Indp
        \ResolveOrdered{$\setOf{\lblVAR}$} = $\emptyset$\;
        \Resolvell{$\lblIMPORT$} = $\setOf{s_2}$\;
        \Shadow{$\emptyset$, $\setOf{s_2}$} = $\setOf{s_2}$\;
        $A$ := $\emptyset \cup \setOf{s_2}$\;
      \Indm
      $\smallerOf{\setOf{\lblVAR, \lblIMPORT, \lblLEX}}{\lblLEX} = \setOf{\lblVAR}$\;
      \Indp
        \ResolveOrdered{$\setOf{\lblVAR}$} = $\emptyset$\;
        \Resolvell{$\lblLEX$} = $\setOf{s_3}$\;
        \Shadow{$\emptyset$, $\setOf{s_3}$} = $\setOf{s_3}$\;
        $A$ := $\setOf{s_2} \cup \setOf{s_3}$ = $\setOf{s_2, s_3}$\;
      \Indm
    \Indm
    \Return $\setOf{s_2, s_3}$\;
  \Indm
\end{boxedalgorithm}

\caption{Execution trace of the algorithm in \cref{fig:edge-traversal-order} applied to $s_E$ in \cref{fig:example-modules}.}
\label{fig:execution-trace}

\end{figure}

To illustrate this, consider the execution trace for the example in \cref{fig:example-modules}, assuming the traversal is in $s_E$, which is shown in \cref{fig:execution-trace}.
Here, the initial $\mathsf{max}$ set is $\setOf{\lblIMPORT, \lblLEX}$ (line 3).
The algorithm iterates over these labels in lines 4 -- 8 ($\pathlbl = \lblIMPORT$) and 9 -- 12 ($\pathlbl = \lblLEX$).
Each of these iterations computes $\setOf{\lblVAR}$ as its $\mathsf{smaller}$ set, and by recursively applying \ResolveOrdered finds that its corresponding environment is empty.
Then the $\pathlbl$ environment is computed, shadowed relative to the \lblVAR-environment (which leaves it unchanged), and added to $\Answer$.
The result of the second iteration is unified with the result of the first iteration, and then returned (line 13 and 14).
This shows how both \lblIMPORT and \lblLEX are shadowed by \lblVAR, but not by each other.

\paragraph{Data well-formedness}
Finally, we select valid declarations using the data well-formedness condition $\wfd$.
This is simply done by evaluating $\wfd$ on the datum of the current scope.
However, there is a small subtlety that must be accounted for.
The resolution algorithm may visit scopes via a path that does not match the original $\re$, but is only a \emph{prefix} of a sequence of labels matched by $\re$.
We should traverse these scopes, but not return them as declaration.
Full paths are precisely those on which the language of the derivative at the targets scope $\re'$ (written as $\langOf{\re'}$) includes the empty word $\reeps$.
Only full paths are considered by $\wfd$ to be included in the query answer.

Returning to the example in \cref{fig:example-modules} again, this means that $\wfd$ is not applied to the datum of $s_C$ (among other scopes), as the language of the current RE ($\lblVAR$) does not include $\reeps$.
However, after traversing the edge to $s_2$, the RE becomes $\derive{\lblVAR} \lblVAR = \reeps$, which obviously matches $\reeps$.
Therefore, $\wfd$ is applied to the datum of $s_2$, selecting it as a valid declaration.

\subsection{Performance of the Resolution Algorithm}
\label{subsec:resolution-performance}

Recall from the introduction that query resolution is slow.
Analyzing the algorithm outline can give an intuition why that is the case.
First, observe that computing the $\mathsf{max}$ and $\mathsf{smaller}$ sets, which is executed per scope the resolution algorithm traverses, is quadratic in the number of labels.
Similarly, the number of derivatives computed per scope traversal is linear in the number of labels.
Therefore, for large scope graphs and label sets, the overhead of these computations can be significant.
Profiling the Java Commons IO project shows that almost 50\% of the total query resolution time is spent on computing derivatives and checking label orders.
Therefore, we lift these computations to specification compile-time, ensuring faster execution.

\begin{figure*}
  \begin{boxedalgorithm}
    \Fn{\Resolve{$\SG, \scope, \re, \wfd, \lblOrd, \equivd$} \withRT\AnswerSets}{
      \vspace{\localfundefskip}
      \LocalFn{\ResolveEOP{$\sgpath$} \withRTNR\AnswerSets}{%
        \Ternary{$\wfd(\assocOf{\target{\sgpath}})$}{$\setOf{ \sgpath }$}{$\emptyset$}%
      }%
      \LocalFn{\Resolvel{$\sgpath, \edgelbl, \re'$} \withRTNR\AnswerSets}{%
        $\mathrel{\bigcup} \setOf{ \ResolveAll{$\pathseg{\sgpath}{\edgelbl}{s'}, \derive{l}\re'$} \alt \edge{\target{\sgpath}}{\edgelbl}{\scope'} \in \edges_{\SG}, \scope' \notin \sgpath}$
      }%
      \EndLocalFnDef
      \LocalFn{\Resolvell{$\sgpath, \pathlbl, \re'$} \withRTNR\AnswerSets}{%
        \Ternary{$\pathlbl = \EOP$}{\ResolveEOP{$\sgpath$}}{\Resolvel{$\sgpath, \pathlbl, \re'$}}%
      }
      \LocalFn{\Shadow{$\Answer_L, \Answer_l$} \withRTNR\AnswerSets}{%
        $\Answer_L \cup \setOf{ \sgpath \in \Answer_l \alt \not\exists \sgpath' \in \Answer_L.\; \equivdOf{\assocOf{\target{\sgpath'}}}{\assocOf{\target{\sgpath}}}}$
      }%
      \EndLocalFnDef
      \LocalFn{\ResolveLl{$\sgpath, \pathlbls, \pathlbl, \re'$} \withRTNR\AnswerSets}{%
        \Shadow{\ResolveL{$\sgpath, \pathlbls, \re'$}, \Resolvell{$p, \pathlbl, \re'$}}
      }%
      \EndLocalFnDef
      \LocalFn{\ResolveL{$\sgpath, \pathlbls, \re'$} \withRTNR\AnswerSets}{%
        $\mathrel{\bigcup} \setOf{ \ResolveLl{$p, \pathlbls', \pathlbl, \re'$} \alt \pathlbl \in \maxOf{\pathlbls}, \pathlbls' = \smallerOf{\pathlbls}{\pathlbl} }$
      }%
      \EndLocalFnDef
      \LocalFn{\ResolveAll{$\sgpath,\re'$} \withRTNR\AnswerSets}{%
        \ResolveL{$\sgpath, \rehead{\re'} \cup \setOf{ \EOP \alt \reeps \in \langOf{\re'}}, \re'$}
      }%
      \EndLocalFnDef
      \Return \ResolveAll{$\scope, \re$}
    }

    \vspace{-0.3em}
  \end{boxedalgorithm}

  \caption{Query Resolution Algorithm (adapted from \Citenamed[alg. 5]{AntwerpenV21})}
  \label{fig:query-resolution-algorithm}
\end{figure*}


\subsection{The Resolution Algorithm}
\label{subsec:resolution-algorithm}

In order to understand the specialization, we first need to understand the generic query resolution algorithm in full detail.
Therefore, we present this algorithm now.

For the algorithm, we use the following notation.
Scope graphs $\SG$ are defined as a three-tuple $\SGrep{\scopes \subset \Scopes}{\edges \subset \Edges}{\dataassoc}$ of scopes $\scope$, edges $e$ and a mapping from scopes to data $d \in \Data$.
Its components can be projected by using $\SG$ as subscript (e.g., $\scopesOf{\SG}$ refers to the scopes of $\SG$).
An edge $\edge{\scope}{\edgelbl}{\scope'}$ consists of a source $s$, a target $s'$ and an edge label $l \in \Labels$.
We define path labels $\pathlbl$ to be either an edge label or the end-of-path label $\EOP$.
$\labels$ denotes a set of edge labels not containing $\EOP$, and $\pathlbls$ is set of labels that might contain $\EOP$.
A path $\sgpath \in \Paths$ can be a single scope $\scope$ or a path step $\pathseg{\sgpath}{\edgelbl}{\scope}$.
The source and target of a path can be projected using $\source{\cdot}$ and $\target{\cdot}$.
An environment $\Answer \in \AnswerSets$ is a set of resolution paths.

The full algorithm is shown in \cref{fig:query-resolution-algorithm}.
It has six inputs: the scope graph, the scope in which to start the query, and the three query parameters.
The last argument, called the \emph{data equivalence condition} ($\equivd$) is a binary predicate over datums.
This argument is used for advanced shadowing, such as shadowing based on overload resolution.
Discussing this argument is out of scope for this paper.
For a full treatment, we refer to \Citenamed{AntwerpenPRV18} and \citenamed{RouvoetAPKV20}.

The algorithm definition uses a number of inner functions that each have their own arguments.
In particular, each function receives a path $\sgpath$, which is the path that the query resolution algorithm traversed so far.
Moreover, most functions have an $\re'$ argument, which is the derivative of $\re$ with respect to the labels in $\sgpath$.
The algorithm starts by computing the full environment of the initial scope and $\re$ using \ResolveAll (line 9).
This function computes the set of labels the regular expression can follow ($\rehead{\re'}$), and includes $\EOP$ when the empty word $\reeps$ is in the language described by the regular expression ($\langOf{\re'}$).
As we well see, including $\EOP$ ensures the local scope is checked for being a valid declaration.

The actual, correctly shadowed environment for these labels is computed using \ResolveL and \ResolveLl.
In fact, \ResolveL resembles \ResolveOrdered from \cref{subsec:algorithm-outline}, where \ResolveLl is the body of its \textbf{for}-loop.
\ResolveL computes the $\mathsf{max}$ and $\mathsf{smaller}$ sets of its $\pathlbls$ argument.
Their respective environments are determined by \ResolveLl using \ResolveL (for the $\mathsf{smaller}$ set $\pathlbls$) and \Resolvell (for the current $\mathsf{max}$ label $\pathlbl$).
These environments are passed to \Shadow, which returns the union of $\Answer_L$ and the elements of $\Answer_l$ that are not shadowed by an element in $\Answer_L$.
That is, elements from $\Answer_l$ that are shadowed by an element from $\Answer_L$ are removed from the final environment.

Computing an environment for a single label is done using the \Resolvell function.
When the $\pathlbl$ is the end-of-path label $\EOP$, the current path is included in the result when its datum matches $\wfd$ (\ResolveEOP).
For an edge label $\edgelbl$, \Resolvel retrieves all outgoing edges of the target of the current path.
For each edge, \ResolveAll is invoked.
The given arguments are the current path, extended with a step that represents the traversed edge, and the derivative of the current regular expression $\re'$ with respect to the current label.
This retains the invariant mentioned above, which ensures that paths returned by the algorithm are valid according to $\re$.

Finally, we want to highlight a few characteristics of the algorithm.
First, note that this algorithm does not enforce local scopes to be considered first.
For example, a query that has label order $\lblL \lblLE \EOP$ will traverse $\lblL$ before checking the local scope.
Second, an invocation of \ResolveAll after traversing an edge in \Resolvel can very well be seen as an independent query.
We refer to such queries as \emph{residual}.

\section{An Intermediate Resolution Language}
\label{sec:query-resolution-language}

Before we define the specializer for queries, we specify the language in which the output is represented.
To this end, we introduce a new intermediate language, which we present in this section.

\subsection{Syntax}
\label{subsec:syntax}

The language that allows us to express specialized versions of name resolution queries is shown in \cref{fig:resolution-language-syntax}.
First, the syntax of query parameters as discussed in the previous section is formalized.
Then, there are names for variables ($\resVar$) and states ($\stateId$).
Variables can be used in \emph{resolution expressions} ($\resExp$).
There are four possible expressions:
\kwResolve resolves the current path;
\kwSubenv traverses all edges with label $\edgelbl$, executing residual queries in state $\stateId$;
\kwMerge merges a collection of environments; and
\kwShadow computes the union of two environments, filtering shadowed paths from the second environment.
A sequence of assignments of expressions to variables constitutes a state ($\State$).
A state machine ($\StateMachine$) consists of a sequence of states, which each are identified by a name.
Implicitly, its first state is designated as initial state.
Finally, the figure shows how this language is embedded in Statix.
There is the traditional variant of Statix ($C$) \cite{AntwerpenPRV18, RouvoetAPKV20}, that has a generic query constraint, which can be resolved using the resolution algorithm introduced in the previous section.
In this paper, we define $C^\ast$, which is a variation of $C$ with the generic query removed and a compiled query added.
This compiled query does not have a path well-formedness condition nor a label order as arguments, but a state machine instead.
After discussing the semantics of query resolution using state machines, we present a compilation scheme from $C$ to $C^\ast$ in \cref{sec:compiling-declarative-queries}.

\begin{figure}
  \begin{adjustbox}{minipage={\linewidth}, padding=1ex, frame, max width = \linewidth}
    \figuresection{Query Parameters}
    \begin{align*}
      \renewcommand\arraystretch{1.0}
      \begin{array}{r O c O l T}
        \re       & \in  & \RE       & \subset & \Labels^\ast                     & \hspace{1.3em} Path well-formedness condition \\
        \wfd      & \in  & \Wfd      & \subset & \Data                            & Data well-formedness predicate \\
        \lblOrd   & \in  & \LblOrd   & \subset & \PathLabels \times \PathLabels   & Label order \\
        \equivd   & \in  & \Equivd   & \subset & \Data \times \Data               & Data equivalence condition \\
      \end{array}
    \end{align*}
    \vspace{-0.5em}
    \figuresection{Resolution State Machine}
    \begin{align*}
      \renewcommand\arraystretch{1.0}
      \begin{array}{r O c O l T}
        \resVar       & \in  & \ResVars       &      &                                                                          & Variables \\
        \stateId      & \in  & \StateIds      &      &                                                                          & State Names \\
        \resExp       & \in  & \ResExps       & ::=  & \expResolve \alt \expSubenv{\edgelbl}{\stateId} \\
                      &      &                & \alt & \expMerge{\arrayOf{\resVar}} \alt \expShadow{\resVar}{\resVar}           & Expressions \\
        \State        & \in  & \States        & ::=  & \expState{\arrayOf{\resVar := \resExp}}{\resVar}                         & States \\
        \StateMachine & \in  & \StateMachines & ::=  & \expStateMachine{\arrayOf{\stateId{:}\; \State}}{\stateId}               & \hspace{1.975em} State Machine
      \end{array}
    \end{align*}
    \vspace{-0.5em}
    \figuresection{Statix}
    \begin{align*}
      \renewcommand\arraystretch{1.0}
      \begin{array}{r O l T}
        C      & ::= & \ldots \alt \qrep{\re}{\wfd}{\lblOrd}{\equivd}{\scope}{\resVar}         & Generic Statix \\
        C^\ast & ::= & \ldots \alt \cqrep{\StateMachines}{\wfd}{\equivd}{\scope}{\resVar}      & Compiled Statix
      \end{array}
    \end{align*}
  \end{adjustbox}
  
  \caption{Syntax of Query Resolution Language}
  \label{fig:resolution-language-syntax}  
  
\end{figure}

\subsection{Semantics}

In this section, we explain how queries in our intermediate language are interpreted (see in \cref{fig:resolution-language-semantics}).
Where appropriate, we explain how it relates to the algorithm in \cref{fig:query-resolution-algorithm}.
We use the following notation.
First, $\StateMachine(\cdot): \StateIds \rightharpoonup \States$ retrieves the state with a particular name from $\StateMachine$.
Second, we use $\sminit{\cdot} : \StateMachines \to \States$ to retrieve the initial state of a state machine.
Finally, $\envstore$ represents a \emph{store} that maps variables ($\resVar$) to environments ($\Answer$).
$\epsilon$ represents the empty store, $\envstore; (\resVar, \Answer)$ denotes adding a mapping $\resVar \mapsto \Answer$ to $\envstore$, and $\envvalue{\cdot} : \ResVars \rightharpoonup \AnswerSets$ retrieves the value (an environment) of a variable.

\begin{figure*}
  \begin{adjustbox}{minipage={\linewidth}, padding=1ex, frame, max width = \linewidth}
	  \setpremisesend{0.5ex}
    \addtolength{\jot}{0.45em}
    \figuresection[\fbox{$\expsem[\envstore][\StateMachines][\SG][\Paths][\Wfd][\Equivd\!\!]{\ResExps}{\AnswerSets}$}]{Expression Evaluation Semantics}
    \begin{gather*}
      \inference[\rulename{\ruleExpResolve}]{
        \Answer = \setOf{p \alt \wfd(\assocOf{\target{p}})}
      }{
        \expsem[\envstore][\StateMachine][\SG][\sgpath][\wfd][\equivd]{\kwResolve}{\Answer}
      }
      \qquad
      \inference[\ruleExpSubenv]{
        \State = \StateMachine(\stateId)
        \qquad
        P = \setOf{\pathseg{\sgpath}{\edgelbl}{\scope'} \alt \edge{\target{\sgpath}}{\edgelbl}{\scope'} \in \edgesOf{\SG}}
        \\
        \Answer = \bigcup\: \setOf{\Answer' \alt \sgpath' \in P,\, \statesem[\StateMachine][\SG][\sgpath'][\wfd][\equivd]{\State}{\Answer'}}
      }{
        \expsem{\expSubenv{\edgelbl}{\stateId}}{\Answer}
      }
      \\
      \inference[\ruleExpMerge]{
        \Answer = \bigcup\: \setOf{\envvalue{\resVar} \alt \resVar \in \arrayOf{\resVar}}
      }{
        \expsem{\expMerge{\arrayOf{\resVar}}}{\Answer}
      }
      \qquad
      \inference[\ruleExpShadow]{
        \Answer_1 = \envvalue{\resVar_1} \qquad \Answer_2 = \envvalue{\resVar_2}
        \\
        \Answer = \Answer_1 \cup \setOf{\sgpath_2 \in \Answer_2 \alt \not\exists \sgpath_1 \in \Answer_1{.}\, \equivdOf{\assocOf{\target{\sgpath_1}}}{\assocOf{\target{\sgpath_2}}}}
      }{
        \expsem[\envstore][\StateMachine][\SG][\sgpath][\wfd][\equivd]{\expShadow{\resVar_1}{\resVar_2}}{\Answer}
      }
    \end{gather*}
    \figuresection[\fbox{$\statesem[\StateMachines][\SG][\Paths][\Wfd][\Equivd\!\!]{\States}{\AnswerSets}$}]{State Evaluation Semantics}
    \begin{gather*}
      \inference[\ruleEvalState]{
        \envstore_0 = \epsilon
        \quad
        \forall_{i \in 1 \ldots n}{.}\, \expsem[\envstore_{i-1}][\StateMachine][\SG][\sgpath][\wfd][\equivd]{\resExp_i}{\Answer_i} \ \ \envstore_i = \envstore_{i-1}{;}(\resVar_i, \Answer_i)
      }{
        \statesem{\expState{\arrayOf{\resVar_i := \resExp_i}_{i = 1 \ldots n}}{\resVar}}{\Answer_n}
      }
    \end{gather*}
    \figuresection[\fbox{$\langle \SG \alt \arrayOf{C^\ast} \rangle \rightarrow \langle \SG \alt \arrayOf{C^\ast} \rangle$}]{Compiled Statix Evaluation Semantics}
    \begin{gather*}
      \inference[\ruleResolveQuery]{
        \State = \sminit{\StateMachine}
        \qquad
        \statesem[\StateMachine][\SG][\scope][\wfd][\equivd]{\State}{\Answer}
      }{
        \langle \SG \alt \cqrep{\StateMachine}{\wfd}{\equivd}{\scope}{\resVar}{;}\,\arrayOf{C} \rangle \rightarrow \langle \SG \alt \arrayOf{C}[x/A] \rangle 
      }
    \end{gather*}
  \end{adjustbox}

  \caption{Operational Semantics of Resolution Language}
  \label{fig:resolution-language-semantics}
\end{figure*}

\begin{figure*}
  \input{figures/compilation/example-1}
  \hspace{-1.25ex}%
  \input{figures/compilation/example-2}
  \hspace{-1.25ex}%
  \input{figures/compilation/example-3}

  \caption{
    Examples of Compiled Queries.
    For each example, the top left box contains the path well-formedness expression, the top right box contains the label order, and the bottom box contains the state machine corresponding to those parameters.
  }
  \label{fig:compilation-examples}
\end{figure*}

Following \citenamed{RouvoetAPKV20}, we define the operational semantics of Statix in a small-step style: a solver state that consists of a scope graph and a set of unsolved constraints ($\arrayOf{C^\ast}$) steps to a new state.
Because we added a new construct to the language, we define the \ruleResolveQuery rule to evaluate it.
This rule defines the answer set of the query to be the result of evaluating the initial state with the initial scope as path (similar to line 9 of \cref{fig:query-resolution-algorithm}).
The remaining constraints $\arrayOf{C^\ast}$ in the output state have the environment substituted in free positions of the result variable $\resVar$.
A state is evaluated by sequentially executing all its steps (\ruleEvalState).
Initially, it starts with an empty store, and each step extends the store with a new mapping.
The rule returns the value of the last expression.
\ruleExpResolve defines that a \kwResolve expression returns the current path if its target datum is well-formed according to $\wfd$.
This is similar to the body of \ResolveEOP.
Next, \kwSubenv expressions are evaluated using the \ruleExpSubenv rule.
This rule defines $P$ to be all valid extensions of path $\sgpath$ with label $\edgelbl$ in $\SG$.
For each path in $P$, we evaluate the new state $\State$.
The result is the union of the resolved residual queries.
This is similar to \Resolvel.
However, instead of computing a derivative of a regular expression, we pass a new state.
As our compilation scheme is designed to ensure that a state is equivalent to \ResolveL specialized to its $\re'$ argument, the behavior of this rule matches the \Resolvel function.
The \kwMerge primitive simply does a lookup of all variables in the current store, and returns the union of them.
This operation will be a part of the compilation scheme for specializing \ResolveL.
Finally, the \kwShadow construct resolves its argument environments, and returns the first environment, combined with the second environment with all shadowed paths removed.
This is similar to the \Shadow function.

\section{Specializing Declarative Queries}
\label{sec:compiling-declarative-queries}

Now that we have defined the language in which we can represent our partially specialized queries, we can define the specializer.
We first discuss a few examples, and then present its full definition.

\subsection{Examples}
\label{subsec:compilation-examples}

\begin{figure}
  \begin{adjustbox}{minipage={\linewidth}, padding=1ex, frame, max width = \linewidth}
    \vspace{-0.5em}
    \begin{flalign*}
                                 \semSM &: \RE \times \LblOrd \to \StateMachines \\
                 \semSMOf{\re}{\lblOrd} &:= \cquote{\expStateMachine{\cunquote{\arrayOf{n{:}\; Y}}}{???}} \where \\
        \arrayOf{\tuple{n, \re', \sid}} &= \genStatesOf{\re}\\
                     \arrayOf{n{:}\: Y} &= \setOf{\cquote{\cunquote{n}{:}\; \cunquote{\semYOf{\re'}{\lblOrd}{\sid}}} \alt \tuple{n, \re', \sid} \in \arrayOf{\tuple{n, \re', \sid}}}
      %
      \\[0.5em]
                                                         \semY &: \RE \times \LblOrd \times (\sidT) \to \States \\
                                   \semYOf{\re}{\lblOrd}{\sid} &:= \cquote{\expState{\cunquote{\arrayOf{\mathit{\resVar := \resExp}}}}{\resVar}} \where \\
        \tuple{\arrayOf{\mathit{\resVar := \resExp}}, \resVar} &= \semLOf{\rehead{\re} \cup \setOf{\EOP \alt \reeps \in \langOf{\re}}}{\lblOrd}{\sid}
      %
      \\[0.5em]
                                                                    \semL &: \powerset{\PathLabels} \times \LblOrd \times (\sidT) \to \arrayOf{\resVar := \resExp} \times \ResVars \\
                                        \semLOf{\pathlbls}{\lblOrd}{\sid} &:= \tuple{(\StmConcat \arrayOf{\arrayOf{\resVar := \resExp}}) \stmConcat \cquote{\cunquote{\resVar_L} := \expMerge{\cunquote{\arrayOf{\resVar}}}}, \resVar_L} \where \\
                                   \arrayOf{\tuple{\pathlbls', \pathlbl}} &= \setOf{\tuple{\pathlbls', \pathlbl} \alt \pathlbl \in \maxOf{\pathlbls},
                                                                                         \pathlbls' = \smallerOf{\pathlbls}{\pathlbl}} \\
        \tuple{\arrayOf{\arrayOf{\resVar := \resExp}}, \arrayOf{\resVar}} &= \unzipOf{\setOf{ \semLlOf{\pathlbls'}{\pathlbl}{\lblOrd}{\sid} \alt \tuple{\pathlbls', \pathlbl} \in
         \arrayOf{\tuple{\pathlbls', \pathlbl}}}} \\
         \resVar_L &= \freshVar
      %
      \\[0.5em]
                                                 \semLl &: \powerset{\PathLabels} \times \PathLabels \times \LblOrd \times (\sidT) \to \arrayOf{\resVar := \resExp} \times \ResVars \\
           \semLlOf{\pathlbls}{\pathlbl}{\lblOrd}{\sid} &:= \tuple{\arrayOf{\resVar := \resExp}', \resVar_{lL}} \where \\
        \tuple{\arrayOf{\resVar := \resExp}, \resVar_L} &= \semLOf{\pathlbls}{\lblOrd}{\sid}\\
                                              \resExp_l &= \semlOf{\pathlbl}{\sid}\\
                                              \resVar_l &= \freshVar \\
                                           \resVar_{lL} &= \freshVar \\
                          \arrayOf{\resVar := \resExp}' &= \arrayOf{\resVar := \resExp} \stmConcat \cquote{\cunquote{\resVar_l} := \cunquote{\resExp_l}} \stmConcat \cquote{\cunquote{\resVar_{lL}} := \expShadow{\cunquote{\resVar_L}}{\cunquote{\resVar_l}}}
      %
      \\[0.5em]
                        \seml &:  \PathLabels \times (\sidT) \to \ResExps \\
          \semlOf{\EOP}{\sid} &:= \cquote{\expResolve} \\
      \semlOf{\edgelbl}{\sid} &:= \cquote{\expSubenv{\cunquote{\edgelbl}}{\cunquote{\sidOf{\edgelbl}}}}
    \end{flalign*}
  \end{adjustbox}

  \caption{Specializer that translates a path well-formedness condition ($\re$) and a label order ($\lblOrd$) to a state machine ($\StateMachine$).}
  \label{fig:compilation-scheme}
\end{figure}

In \cref{fig:compilation-examples}, a few examples of specialized queries are shown.
The first example shows a query that traverses a single $\lblL$ label and does not apply any shadowing.
Its specialized variant has two states.
State $n_0$ has a single \kwSubenv expression that traverses $\lblL$ labels while transitioning to state $n_1$.
In state $n_1$, the current path is resolved.
So how was this translation performed?
First, each state in the state machine corresponds to a possible derivative of $\re$.
State $n_0$ is derived from the original regex $\lblL$, while state $n_1$ is derived from $\derive{\lblL}\lblL = \reeps$.
Each derivative has a singleton head set ($\setOf{\lblL}$ and $\setOf{\EOP}$, respectively).
Therefore, each state is implemented by resolving that label.
The \kwSubenv expression in $n_0$ transitions to $n_1$ because state $n_1$ corresponds to the derivative with respect to $\lblL$ of the regex of state $n_0$.
In the state machine of the second example, there is only one state, as $\derive{\lblL}\reclos{\lblL} = \reclos{\lblL}$.
However, the head set of that regular expression is $\setOf{\lblL, \EOP}$.
Therefore, state $n_0$ computes both sub-environments, and combines them with the \kwMerge operator, since there is no ordering between the labels.
Compared to \ResolveL, the lack of an ordering means that both labels are in the $\max$-set.
Therefore, both environments are computed by a call to \ResolveLl with an empty $\pathlbls$ set.
This translates to single calls to \ResolveEOP and \Resolvel, of which the union is returned by \ResolveL.
In the third example however, there is an ordering between the labels.
Therefore, the sub-environments are combined using the \kwShadow operator instead.
When we compare this with the execution of \ResolveL again, we see that the $\max$-set is $\setOf{\lblL}$, with a $\mathsf{smaller}$-set of $\setOf{\EOP}$.
In \ResolveLl, this translates to the result of traversing $\lblL$ being shadowed with respect to the local environment, which again corresponds with the behavior of the specialized query.
In summary, these examples show how states correspond with derivatives of regular expressions, and how \kwMerge and \kwShadow are used to model different label orders.

\subsection{Specializer}
\label{subsec:compilation-scheme}

In this section, we present the specializer that generates a state machine $\StateMachine$ for a regular expression $\re$ and a label order $\lblOrd$.
For our presentation, we use the following notation.
First, the $\stmConcat$ infix operator appends an item to a list.
Similarly, its larger variant flattens a list of lists by concatenating its sublists.
In addition, we assume the following helper functions:
\begin{align*}
  \unzipOf{\cdot} &: \forall\, \mathcal{T}_1\, \mathcal{T}_2.\; \arrayOf{\mathcal{T}_1 \times \mathcal{T}_2} \to \arrayOf{\mathcal{T}_1} \times \arrayOf{\mathcal{T}_2} \\
  \freshVar &: () \to \ResVars \\
  \genStatesOf{\cdot} &: \RE \to \arrayOf{\StateIds \times \RE \times \Labels \rightharpoonup \StateIds}
\end{align*}
The $\unzip$ function translates a list of 2-tuples in a tuple of lists.
Next, the $\fv$ function generates a fresh variable at each invocation.
Finally, the $\genStates$ function expands a regular expression into its state machine.
This state machine is defined as an array of three-tuples, where each tuple (a state) contains
(1) a unique name,
(2) the derivative of the original regular expression that corresponds to current state, and
(3) a partial transition function (defined for the head set of (2)) that maps labels to the identifier of the state.
The result of $\genStates$ has two invariants.
First, the first state in the list should be the initial state.
That is, its second component should be equal to the original input of the function.
Second, given two states $\tuple{n, \re, \sid}$ and $\tuple{n', \re', \sid'}$, and a label $\edgelbl$ such that $\sidOf{\edgelbl} = n'$, then $\derive{\edgelbl} \re = \re'$.
This invariant ensures the transitions in the state machine correspond to the original regular expression.
  We implemented $\genStates$ as follows.
  We construct a DFA for the input regular expression~\cite{Sipser12}, assign each node in the DFA a name, and then construct an entry for each node.
  Each entry contains the generated name, the regular expression corresponding to the node \cite{Brzozowski64, OwensRT09} and a transition function based on the transitions in the DFA.

\Cref{fig:compilation-scheme} shows $\semSM$, which specializes a regular expression and a label order into a state machine.
In $\semSM$, $\genStates$ generates a DFA, which is subsequently compiled using $\semName{\State}$ on each state.
This function generates a sequence of expressions that computes an environment for each label in the head set of the regular expression argument, including the $\EOP$ label if the regular expression matches the empty word (similar to \ResolveAll).
The code for this environment is generated by the $\semName{\mathsf{L}}$ function, which, similar to \ResolveL, first computes the $\max$-set of its $\pathlbls$ argument.
For each label in this set, code that computes its shadowed environment is generated.
This yields an array of statement sequences and an array of variables.
The first array is flattened to obtain a sequence of statements that computes all sub-environments.
At the end, a $\kwMerge$ operation is appended that stores the merged environment in the fresh $\resVar_L$ variable.
This sequence is returned, together with $\resVar_L$, which ensures callees can refer to the new environment.
Code to compute a shadowed environment is generated using the $\semName{\mathsf{lL}}$ function.
This function generates code for the environment of its $\pathlbls$ argument, and an expression that creates the environment of its $\pathlbl$ argument.
Then, it appends statements that store the latter in the fresh $\resVar_l$ variable, and subsequently create a shadowed environment.
This sequence is returned along with the result variable $\resVar_{lL}$.
Finally, code that resolves single-label environments is generated by $\semName{\mathsf{l}}$.
For the $\EOP$ label, $\kwResolve$ is returned, while for an edge label $\edgelbl$, a $\kwSubenv$ construct is used.
The $\sid$ function is used to find the state the query resolution needs to transition to when traversing $\edgelbl$ edges.
The second invariant on $\genStates$ ensures that $\sidOf{\edgelbl}$ is the state that implements \ResolveAll specialized to $\derive{\edgelbl} \re$.
This ensures the behavior of this expression is equal to the \Resolvel function.

\begin{figure}
\begin{adjustbox}{minipage={\linewidth}, padding=0ex 0ex, frame}%
\begin{adjustbox}{minipage={0.45\linewidth}, padding=1ex 0ex}%
\begin{lstlisting}[language=SM]
n:
  e$\cidx{0}$ := subenv @lbl@L$\cidx{1}$@ n
  e$\cidx{1}$ := subenv @lbl@L$\cidx{2}$@ n
  e$\cidx{2}$ := shadow e$\cidx{0}$ e$\cidx{1}$
  e$\cidx{3}$ := subenv @lbl@L$\cidx{1}$@ n
  e$\cidx{4}$ := subenv @lbl@L$\cidx{3}$@ n
  e$\cidx{5}$ := shadow e$\cidx{3}$ e$\cidx{4}$
  e$\cidx{6}$ := merge e$\cidx{2}$ e$\cidx{5}$
\end{lstlisting}%
\end{adjustbox}%
\vrule%
\begin{adjustbox}{minipage={0.55\linewidth}, padding=1ex 0ex}%
\begin{lstlisting}[language=SM]
n:
  e$\cidx{0}$ := subenv @lbl@L$\cidx{1}$@ n
  e$\cidx{1}$ := subenv @lbl@L$\cidx{2}$@ n
  e$\cidx{2}$ := shadow e$\cidx{0}$ e$\cidx{1}$
  e$\cidx{3}$ := subenv @lbl@L$\cidx{3}$@ n
  e$\cidx{4}$ := shadow e$\cidx{0}$ e$\cidx{3}$
  e$\cidx{5}$ := merge e$\cidx{2}$ e$\cidx{4}$
@lbl@@
\end{lstlisting}%
\end{adjustbox}
\end{adjustbox}

\caption{Example of optimization based on available expressions.}
\label{fig:available-expressions-example}

\end{figure}

\subsection{Eliminating Common Sub-Environments}
\label{subsec:available-subenvs}

Having an intermediate language opens up some additional opportunities for optimization as well.
First, consider a case where the label order is defined as $\lbldef{L_1} \lblLE \lbldef{L_2}, \lbldef{L_1} \lblLE \lbldef{L_3}$.
Applying the specializer yields the result from the left side of \cref{fig:available-expressions-example}.
However, the values of \lstinline{e$\cidx{0}$} and \lstinline{e$\cidx{3}$} are equal.
Therefore, we can eliminate the calculation of \lstinline{e$\cidx{3}$}, and simply use \lstinline{e$\cidx{0}$} instead.
In general, this corresponds to applying \emph{common sub-expression elimination}.
As this can save redundant computation of sub-environments, this optimization can have a significant impact on the total run time.

\subsection{Skipping Fully Shadowed Environments}
\label{subsec:skipping-shadowed-environments}

For the second optimization, recall that \kwShadow evaluates to
\[
\Answer_1 \cup \setOf{\sgpath_2 \in \Answer_2 \alt \not\exists \sgpath_1 \in \Answer_1{.}\, \equivdOf{\assocOf{\target{\sgpath_1}}}{\assocOf{\target{\sgpath_2}}}}
\]
where $\Answer_1$ is the environment that can shadow declarations in $\Answer_2$.
Now consider the case that all declarations can shadow each other (i.e., $\forall\, d_1\mkern1mu d_2.\, \equivdOf{d_1}{d_2}$).
As we have seen in \cref{sec:query-resolution-in-scope-graphs}, that is the most common situation.
In this case, any element in $\Answer_1$ will shadow $\Answer_2$ completely.
Thus, the \kwShadow operator can be simplified to choosing $\Answer_1$ if it is not empty, and $\Answer_2$ otherwise.
However, that means that we do not need to compute $\Answer_2$ at all when $\Answer_1$ is not empty.

This optimization is implemented using a small extension of the query resolution language.
An \kwElse operator, as shown in \cref{fig:resolution-language-extension}, is added, which has a variable and a sub-expression as operands.
When the environment of the variable is non-empty, it is returned ($\ruleExpElseL$).
In this case, the right-hand expression is not evaluated.
When the variable is empty, that expression is evaluated and its value returned ($\ruleExpElseR$).

\begin{figure}
  \begin{adjustbox}{minipage={\linewidth}, padding=0ex, frame}
  \begin{adjustbox}{minipage={\linewidth}, padding=0ex}
    \vspace{-0.5em}
    \begin{align*}
      \renewcommand\arraystretch{1.0}
      \begin{array}{r O c O l T}
        \resExp       & \in  & \ResExps       & ::=  & \ldots \alt \expElse{\resVar}{\resExp}  & Expressions \\
      \end{array}
    \end{align*}
  \end{adjustbox}
  \hrule
  \begin{adjustbox}{minipage={\linewidth}, padding=1ex}
    \begin{gather*}
      \inference[\ruleExpElseL]{
        \envvalue{\resVar} = \Answer
        \qquad
        \Answer \neq \emptyset
      }{
        \expsem[\envstore][\StateMachine][\SG][\sgpath][\wfd][\equivd]{\expElse{\resVar}{\resExp}}{\Answer}
      }
      \\
      \inference[\ruleExpElseR]{
        \envvalue{\resVar} = \emptyset
        \qquad
        \expsem[\envstore][\StateMachine][\SG][\sgpath][\wfd][\equivd]{\resExp}{\Answer}
      }{
        \expsem[\envstore][\StateMachine][\SG][\sgpath][\wfd][\equivd]{\expElse{\resVar}{\resExp}}{\Answer}
      }
    \end{gather*}
  \end{adjustbox}
  \hrule
  \begin{adjustbox}{minipage={\linewidth}, padding=0ex}
  \end{adjustbox}
    \begin{flalign*}
           \semLlOf{\pathlbls}{\pathlbl}{\lblOrd}{\sid} &:= \tuple{\arrayOf{\resVar := \resExp}', \resVar_{lL}} \where \\
        \tuple{\arrayOf{\resVar := \resExp}, \resVar_L} &=  \semLOf{\pathlbls}{\lblOrd}{\sid}\\
                                              \resExp_l &=  \semlOf{\pathlbl}{\sid}\\
                                           \resVar_{lL} &=  \freshVar \\
                          \arrayOf{\resVar := \resExp}' &=  \arrayOf{\resVar := \resExp} \stmConcat \cquote{\cunquote{\resVar_{lL}} := \expElse{\cunquote{\resVar_L}}{\cunquote{\resExp_l}}}
    \end{flalign*}
  \end{adjustbox}

  \caption{Resolution Language Extension}
  \label{fig:resolution-language-extension}

\end{figure}

The bottom part of \cref{fig:resolution-language-extension} shows an alternative version of $\semLl$ that uses this expression.
In our specializer, this version is used when $\equivd$ is trivially satisfied.
Similar to the other version, it generates code that computes the environment of $\pathlbls$, an expression that computes the environment of $\pathlbl$ ($\resExp_{\pathlbl}$), and a result variable $\resVar_{lL}$.
To the sequence of code, it appends a statement that stores either $\resVar_{L}$ or $\resExp_{\pathlbl}$ in $\resVar_{lL}$.
This sequence of code and the result variable $\resVar_{lL}$ is then returned.

Our implementation of the query resolution algorithm contained this optimization as well.
However, lifting it to specification compile-time removed the overhead of checking whether the data equivalence condition is trivially true.

\section{Evaluation}
\label{sec:evaluation}

To evaluate the correctness and performance of our optimization, we applied a Statix specification for a subset of Java with pre-compiled queries on the Apache Commons CSV, IO and Lang3 projects.
We used these projects as evaluation corpus because they have scope graphs of significant size, and use queries with complex path well-formedness conditions and label orders.
Thus, for these projects performance problems are most urgent.
All benchmarks are executed with the JMH benchmark tool~\cite{JMH}, on a Linux system with 2 AMD EPYC 7502 32-Core Processors (1.5GHz, 2 threads) and 256GB RAM.
\Cref{fig:benchmark-summary} shows a summary of the benchmark results.

\subsection{Correctness}
\label{subsec:eval-correctness}

Essential for the validity of an optimization is its correctness.
An optimization that produces incorrect results is of no value.
This especially holds for Statix, where soundness with respect to its declarative semantics is essential~\cite{AntwerpenPRV18, RouvoetAPKV20}.
Therefore, we validated that for each of the 253177 queries executed by the evaluation projects, the generic algorithm yielded the same answer as the specialized version.
As this was the case for a collection of complex queries in large scope graphs, we have enough confidence that our approach is correct.

\subsection{Performance}
\label{subsec:eval-performance}

\begin{figure}

  \centering

  \vspace{-1.5em}
  \includegraphics[width=\linewidth]{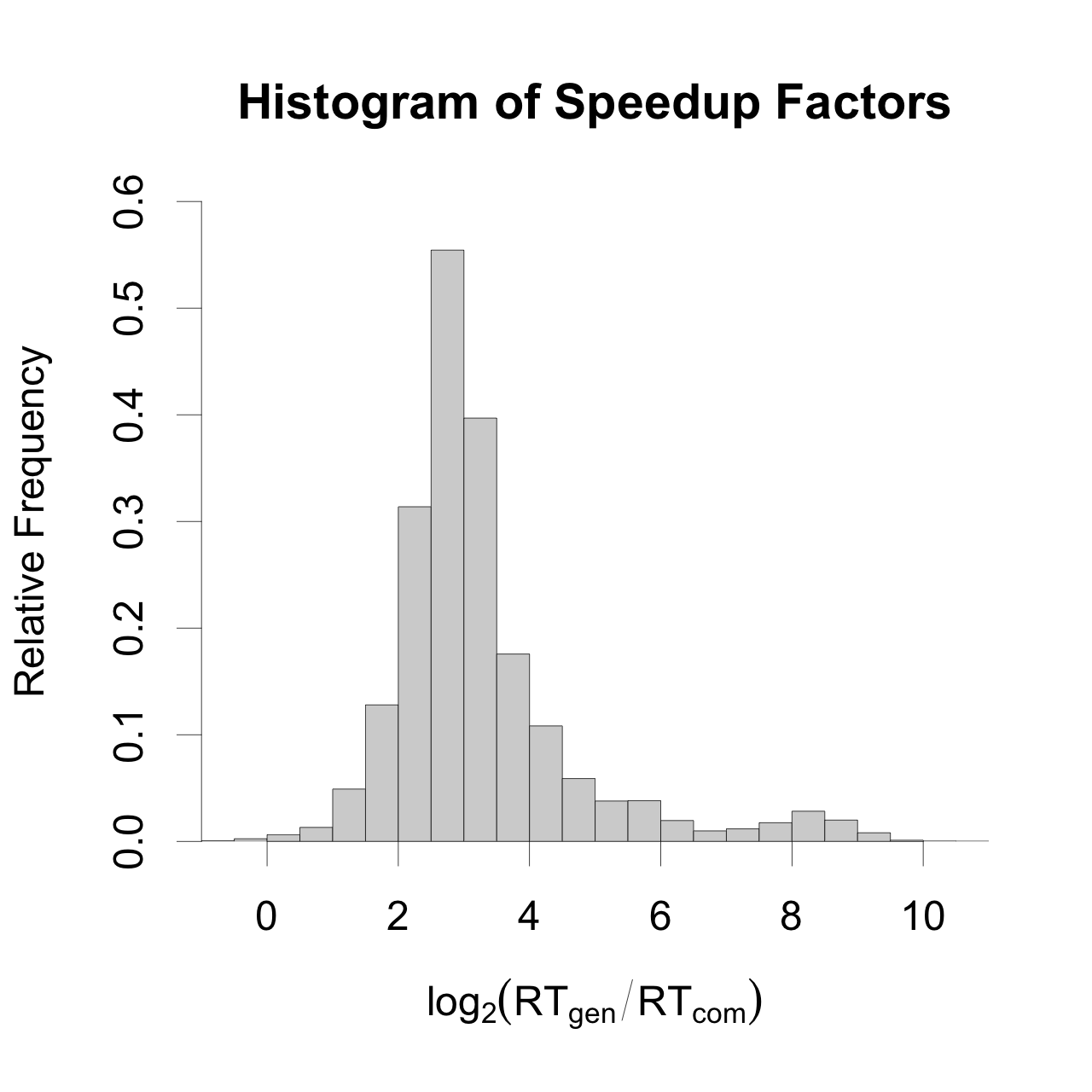}
  \vspace{-1.5em}

  \begin{tabular}{c c c c c | c}
    \toprule
    \textit{Min} & \textit{Q1} & \textit{Q2} & \textit{Q3} & \textit{Max} & Upper fence (\textit{Q2} + 1.5$\mathit{IQR}$) \\
    \midrule
    0.52         & 5.61        & 7.7         & 12.9         & 1985        & 17.2                                          \\
    \bottomrule
  \end{tabular}

  \caption{
    Histogram and five-number summary with upper fence of individual query speedup factors.
    The x-axis of the histogram displays speedup factors in logarithmic scale.
    In the table, \textit{Q1} -- \textit{Q3} represent the quartiles.
    $\mathit{IQR}$ is the inter-quartile range (\textit{Q3} - \textit{Q1}).
  }
  \label{fig:speedup-plot}

\end{figure}

In addition, we considered how much our optimization improved performance of query resolution.
To this end, we traced all individual queries of the CSV project.
For each individual query, we separately measured the performance of the generic and the compiled version.
This benchmark used 8 warmup iterations of 500ms, and 5 measurement iterations of 2000ms with 4 parallel threads in throughput mode.
A summary of the results is shown in \cref{fig:speedup-plot}.
As the speedup distribution is skewed left, but has a relatively heavy tail, the histogram is plot in log-scale, and the upper fence is shown.
The individual speedups range from 0.52 (a slowdown) to 1985, with the most weight around the mean of 7.7.
This shows that our approach was able to lift much of the computation to specification compile time.

\begin{figure}
  \centering
  \begin{tabular}{l | c | c c c}
    \toprule
    \textbf{Project} & \textbf{\#Queries} & $\mathbf{RT_{gen}}$ & $\mathbf{RT_{com}}$ & \textbf{Speedup} \\
    \midrule
      CSV 1.7        &  10599             & 7.3                 & 4.5                 & 39\%             \\
      IO 2.6         &  56413             & 19                  & 12                  & 38\%             \\
      Lang3 3.11     & 186165             & 88                  & 46                  & 48\%             \\
    \bottomrule
  \end{tabular}
  \caption{
    Benchmark Summary.
    The third and fourth column give the total run times in seconds when using the generic algorithm versus compiled queries, respectively.}
  \label{fig:benchmark-summary}
\end{figure}

Executing a Statix specification entails more than executing queries.
Therefore, these results cannot be interpreted as speedups for Statix-based type checkers.
For that reason, we benchmarked the total speedup of executing the Java specification with the concurrent solver~\cite{AntwerpenV21}, using both generic and compiled queries. 
For each measurement, we used 5 warm-up iterations and 20 measurement iterations in single-shot mode on a single core.
The results are shown in \cref{fig:benchmark-summary} as well.
We see that the overall performance of the Java specification improved with 38 -- 48\%.
This is partially due to the elimination of computing label orders and regular expression derivatives, but also to the new optimization discussed in \cref{subsec:available-subenvs}, which our intermediate language allowed to implement.
Although this does not attain performance comparable to \texttt{javac}, it is a significant step in that direction.

\subsection{Compilation}
\label{subsec:compilation}

In addition, we need to assess the compile-time overhead query specialization introduces.
Thus, we measured the time required for specializing queries for the Java specification.
We found that it is responsible for 4.6\% of the total compilation time, which is an acceptable overhead.

\subsection{Threats to Validity}
\label{subsec:threats-to-validity}

There are three threats to the validity of the evaluation that we discuss now.
First, we did not explicitly verify the coverage of our correctness validation.
Although we deem it unlikely, it might be the case that queries of the Java specification do not exercise particular important code paths.

Second, Statix allows interleaving scope graph construction and querying~\cite{RouvoetAPKV20, AntwerpenV21}.
This is required to support type-dependent name resolution and module systems.
To ensure that query answers are valid, some internal scheduling is done.
In addition, evaluating data well-formedness conditions can be delayed when a datum contains free unification variables.
The benchmarks for individual queries were executed on complete scope graphs, where query scheduling and unification was not needed.
Hence, these benchmarks do not account for the overhead that causes.
This threat does not concern the type checker benchmarks, thus the conclusion of a significant speedup on Java programs remains valid.

Third, while our Java benchmark set shows the performance characteristics of the approach clearly, it might not be fully representative.
Statix is often used for language prototyping and DSLs, which often give simpler specifications and smaller codebases.
Still, our evaluation shows that our approach improves the run time of specifications for which performance problems are most relevant.

\section{Related Work}
\label{sec:related-work}

In this section, we discuss related work on scope graphs, declarative type checkers and specializing interpreters.

\subsection{Scope Graphs}

Scope graphs were introduced by \citenamed{NeronTVW15} as a language-parametric model of name binding for languages with non-trivial binding structures.
This model was embedded in NaBL2, which is a DSL for declarative specification of type checkers~\cite{AntwerpenNTVW16}.
NaBL2 employed a strict two-phase approach of constraint generation and solving.
To extend support to structural types and parameterized types, the scope graph model was generalized and embedded in a new DSL: Statix~\cite{AntwerpenPRV18}.
Statix allows the definition of user-defined constraints, which, unlike NaBL2, allows interleaving of constraint solving and introduction of new constraints.
\Citenamed{RouvoetAPKV20} present a formal operational semantics for Statix, and proved it sound with respect to its declarative semantics.
To prove the correctness of query resolution in incomplete scope graphs, the concept of \emph{critical edges} was introduced. 
\Citenamed{AntwerpenV21} generalized this notion to the \emph{scope states} protocol.
Using this protocol, an actor-based concurrent semantics for Statix was presented.

As scope graphs provide uniform program representation model, they have been used to provide language-parametric editor services, such as semantic completion~\cite{PelsmaekerAPV22}, renaming~\cite{Misteli20,Misteli21} and inlining~\cite{Gugten22}.
In addition, it has been shown that scope graphs can be used to describe \emph{frames}, which model the structure of run time heaps \cite{PoulsenNTV16, VerguTV19}.

\subsection{Declarative Type System Specification}

In addition to scope graphs, there is other research into declarative type system specification.
\emph{Attribute Grammars}, such as JastAdd~\cite{HedinM03} and Silver~\cite{WykBGK10} allow the definition of attributes on AST nodes.
Functions that compute attribute values can reference attributes of other nodes, hence abstracting from traditional AST traversal. 
However, language implementers need to define the traversal order manually, whereas scope graph query resolution derives that from declarative queries.
To the best of our knowledge, there is no research about applying partial evaluation to attribute grammar systems.

Although it is more often used for other types of analysis~\cite{SmaragdakisB10, SzaboBEV18, SzaboKBME18}, \emph{datalog} has been used for the specification of type systems as well \cite{PacakES20, ErdwegBKKM15}.
This research was especially aimed at leveraging the good incremental performance of datalog solvers to type checkers.
Datalog-based type systems attain a high level of declarativity and good performance, although scope graphs allow easier encoding of complicated name binding patterns.
\Citenamed{ScholzJSW16} apply partial evaluation of datalog specifications (targeting C++) to gain efficient program analyzers.
This setup is rather similar to ours, although we targeted a tailored intermediate language.

\subsection{Optimizing Interpreters by Partial Evaluation}

Kleene's \textit{s-m-n theorem} \cite{kleene52} essentially proved partial evaluation possible~\cite{Jones:Gomard:Sestoft:93:PartialEvaluation}.
Futamura applied this concept on interpreters, establishing fundamental relations between specializers, interpreters, compilers, source programs and executables \cite{Futamura82}.
The first of his three projections was to specialize an interpreter to a source program, yielding an executable.
In \cref{sec:partial-evaluation}, we generalize over this by introducing partial specialization that changes a program in another (interpreted) program of lower complexity.
Although they do not name it, \Citenamed{ThibaultCLMM00} argue that the first Futamura projection can be applied at compile-time as well as at run time.

\Citenamed{BradyH10} show how to use partial evaluation on DSLs embedded in dependently-typed languages, arguing it is possible to have both correctness and efficiency.
Our paper differs in the fact that our query resolution language is (1) not embedded in a dependently-typed language and (2) has a more complicated interpreter, due to the declarative nature of scope graph queries.
In addition, our specializer targets a tailored intermediate language, instead of its the host language.
This allowed us to carefully consider which parts of the algorithm we specialize, but makes the approach harder to transfer to other languages.

\Citenamed{HumerWWWW14} introduce \textit{Truffle}, which is an embedded DSL for self-optimizing interpreters.
It allows language implementers to annotate specialization possibilities on operations, which are dynamically applied when required.
This allows efficient interpretation of (especially) dynamic languages.
Truffle-based interpreters are often executed on the \textit{Graal VM}~\cite{Graal}, which is an optimizing just-in-time Java compiler.
However, \citenamed{VerguTV19} argue that \emph{meta-interpreting} specifications of dynamic semantics "introduces runtime overhead that is difficult to remove by using interpreter optimization frameworks such as the Truffle/Graal Java tools."
Using scopes and frames however, optimization of meta-interpreters beyond straightforward application of Truffle can be done~\cite{VerguTV19}.
In general, Truffle is particularly aimed at optimization of AST interpreters, which have a very syntax-directed evaluation style.
Whether Truffle provides speedup for more algorithmic interpreters is an interesting question for future research.

\section{Conclusion}
\label{sec:conclusion}

In this paper, we have seen how scope graphs can be used to give high-level encodings of name binding and resolution patterns in programming languages.
In addition, we discussed the algorithm that interprets these declarative queries, yielding actual environments in scope graphs of real programs.
However, this algorithm turned out to impose significant run time overhead.
To eliminate that, 
we applied partial evaluation to Statix, yielding a specializer that translates declarative queries into a more low-level intermediate representation.
These queries can be executed up to 7.7x faster, yielding a speedup of Statix-based type checkers of 38\% -- 48\%.
This is a step toward deriving type checkers from declarative specifications that have performance comparable to hand-written type checkers.

Our work suggests that partial evaluation is a powerful technique to optimize execution of programs written in declarative languages.
Because interpreters of such languages generally perform complex computations on programs, specialization might reduce run time even more for interpreters of more imperative languages.
Further establishing the relation between partial evaluation and interpreters for declarative languages seems a promising topic for further research.

\printbibliography

@article{PacakES20,
  title = {A systematic approach to deriving incremental type checkers},
  author = {André Pacak and Sebastian Erdweg and Tamás Szabó},
  year = {2020},
  doi = {10.1145/3428195},
  url = {https://doi.org/10.1145/3428195},
  researchr = {https://researchr.org/publication/PacakES20},
  cites = {0},
  citedby = {0},
  journal = {Proceedings of the ACM on Programming Languages},
  volume = {4},
  number = {OOPSLA},
}

@inproceedings{ErdwegBKKM15,
  title = {A co-contextual formulation of type rules and its application to incremental type checking},
  author = {Sebastian Erdweg and Oliver Bracevac and  Edlira Kuci and Matthias Krebs and Mira Mezini},
  year = {2015},
  doi = {10.1145/2814270.2814277},
  url = {http://doi.acm.org/10.1145/2814270.2814277},
  researchr = {https://researchr.org/publication/ErdwegBKKM15},
  cites = {0},
  citedby = {0},
  pages = {880-897},
  booktitle = {Proceedings of the 2015 ACM SIGPLAN International Conference on Object-Oriented Programming, Systems, Languages, and Applications, OOPSLA 2015, part of SPLASH 2015, Pittsburgh, PA, USA, October 25-30, 2015},
  editor = {Jonathan Aldrich and Patrick Eugster},
  publisher = {ACM},
  isbn = {978-1-4503-3689-5},
}

@article{SzaboBEV18,
  title = {Incrementalizing lattice-based program analyses in Datalog},
  author = {Tamás Szabó and Gábor Bergmann and Sebastian Erdweg and Markus Voelter},
  year = {2018},
  doi = {10.1145/3276509},
  url = {https://doi.org/10.1145/3276509},
  researchr = {https://researchr.org/publication/SzaboBEV18},
  cites = {0},
  citedby = {0},
  journal = {Proceedings of the ACM on Programming Languages},
  volume = {2},
  number = {OOPSLA},
}

@article{AntwerpenV21-artifact,
  title = {Scope States (Artifact)},
  author = {Hendrik van Antwerpen and Eelco Visser},
  year = {2021},
  doi = {10.4230/DARTS.7.2.1},
  url = {https://doi.org/10.4230/DARTS.7.2.1},
  researchr = {https://researchr.org/publication/AntwerpenV21-artifact},
  cites = {0},
  citedby = {0},
  journal = {DARTS},
  volume = {7},
  number = {2},
}

@inproceedings{AntwerpenV21,
  title = {Scope States: Guarding Safety of Name Resolution in Parallel Type Checkers},
  author = {Hendrik van Antwerpen and Eelco Visser},
  year = {2021},
  doi = {10.4230/LIPIcs.ECOOP.2021.1},
  url = {https://doi.org/10.4230/LIPIcs.ECOOP.2021.1},
  researchr = {https://researchr.org/publication/AntwerpenV21},
  cites = {0},
  citedby = {0},
  booktitle = {35th European Conference on Object-Oriented Programming, ECOOP 2021, July 11-17, 2021, Aarhus, Denmark (Virtual Conference)},
  editor = {Anders Møller and Manu Sridharan},
  volume = {194},
  series = {LIPIcs},
  publisher = {Schloss Dagstuhl - Leibniz-Zentrum für Informatik},
  isbn = {978-3-95977-190-0},
}

@inproceedings{NeronTVW15,
  title = {A Theory of Name Resolution},
  author = {Pierre Néron and Andrew P. Tolmach and Eelco Visser and Guido Wachsmuth},
  year = {2015},
  doi = {10.1007/978-3-662-46669-8_9},
  url = {http://dx.doi.org/10.1007/978-3-662-46669-8_9},
  researchr = {https://researchr.org/publication/NeronTVW15},
  cites = {0},
  citedby = {0},
  pages = {205-231},
  booktitle = {Programming Languages and Systems - 24th European Symposium on Programming, ESOP 2015, Held as Part of the European Joint Conferences on Theory and Practice of Software, ETAPS 2015, London, UK, April 11-18, 2015. Proceedings},
  editor = {Jan Vitek},
  volume = {9032},
  series = {Lecture Notes in Computer Science},
  publisher = {Springer},
  isbn = {978-3-662-46668-1},
}

@article{WykBGK10,
  title = {Silver: An extensible attribute grammar system},
  author = {Eric Van Wyk and Derek Bodin and Jimin Gao and Lijesh Krishnan},
  year = {2010},
  doi = {10.1016/j.scico.2009.07.004},
  url = {http://dx.doi.org/10.1016/j.scico.2009.07.004},
  tags = {attribute grammars, grammar},
  researchr = {https://researchr.org/publication/WykBGK10},
  cites = {0},
  citedby = {0},
  journal = {Science of Computer Programming},
  volume = {75},
  number = {1-2},
  pages = {39-54},
}

@inproceedings{Misteli20,
  title = {Towards language-parametric refactorings},
  author = {Philippe D. Misteli},
  year = {2020},
  doi = {10.1145/3397537.3398476},
  url = {https://doi.org/10.1145/3397537.3398476},
  researchr = {https://researchr.org/publication/Misteli20},
  cites = {0},
  citedby = {0},
  pages = {213-214},
  booktitle = {Programming'20: 4th International Conference on the Art, Science, and Engineering of Programming, Porto, Portugal, March 23-26, 2020},
  editor = {Ademar Aguiar and Shigeru Chiba and Elisa Gonzalez Boix},
  publisher = {ACM},
  isbn = {978-1-4503-7507-8},
}

@inproceedings{ScholzJSW16,
  title = {On fast large-scale program analysis in Datalog},
  author = {Bernhard Scholz and Herbert Jordan and Pavle Subotic and Till Westmann},
  year = {2016},
  doi = {10.1145/2892208.2892226},
  url = {http://doi.acm.org/10.1145/2892208.2892226},
  researchr = {https://researchr.org/publication/ScholzJSW16},
  cites = {0},
  citedby = {0},
  pages = {196-206},
  booktitle = {Proceedings of the 25th International Conference on Compiler Construction, CC 2016, Barcelona, Spain, March 12-18, 2016},
  editor = {Ayal Zaks and Manuel V. Hermenegildo},
  publisher = {ACM},
  isbn = {978-1-4503-4241-4},
}

@book{Sipser12,
  title = {Introduction to the Theory of Computation},
  author = {Michael Sipser},
  year = {2012},
  researchr = {https://researchr.org/publication/Sipser12},
  cites = {0},
  citedby = {0},
  edition = {3rd},
  publisher = {Cengage Learning},
  isbn = {978-1-133-18779-0},
}

@book{Jones:Gomard:Sestoft:93:PartialEvaluation,
  title = {Partial Evaluation and Automatic Program Generation},
  author = {Neil D. Jones and Carsten K. Gomard and Peter Sestoft},
  year = {1993},
  month = {jun},
  note = {ISBN number 0-13-020249-5 (pbk)},
  tags = {partial evaluation},
  researchr = {https://researchr.org/publication/Jones%3AGomard%3ASestoft%3A93%3APartialEvaluation},
  cites = {0},
  citedby = {0},
  editor = {Series editor C. A. R.\ Hoare},
  address = {International Series in Computer Science},
  publisher = {Prentice Hall International},
}

@inproceedings{HumerWWWW14,
  title = {A domain-specific language for building self-optimizing AST interpreters},
  author = {Christian Humer and Christian Wimmer and Christian Wirth and Andreas Wöß and Thomas Würthinger},
  year = {2014},
  doi = {10.1145/2658761.2658776},
  url = {http://doi.acm.org/10.1145/2658761.2658776},
  researchr = {https://researchr.org/publication/HumerWWWW14},
  cites = {0},
  citedby = {0},
  pages = {123-132},
  booktitle = {Generative Programming: Concepts and Experiences, GPCE'14, Vasteras, Sweden, September 15-16, 2014},
  editor = {Ulrik Pagh Schultz and Matthew Flatt},
  publisher = {ACM},
  isbn = {978-1-4503-3161-6},
}

@article{kleene52,
  title = {Introduction to metamathematics},
  author = {Kleene, Stephen Cole},
  year = {1952},
  researchr = {https://researchr.org/publication/kleene52},
  cites = {0},
  citedby = {0},
}

@inproceedings{SmaragdakisB10,
  title = {Using Datalog for Fast and Easy Program Analysis},
  author = {Yannis Smaragdakis and Martin Bravenboer},
  year = {2010},
  doi = {10.1007/978-3-642-24206-9_14},
  url = {http://dx.doi.org/10.1007/978-3-642-24206-9_14},
  researchr = {https://researchr.org/publication/SmaragdakisB10},
  cites = {0},
  citedby = {0},
  pages = {245-251},
  booktitle = {Datalog Reloaded - First International Workshop, Datalog 2010, Oxford, UK, March 16-19, 2010. Revised Selected Papers},
  editor = {Oege de Moor and Georg Gottlob and Tim Furche and Andrew Jon Sellers},
  volume = {6702},
  series = {Lecture Notes in Computer Science},
  publisher = {Springer},
  isbn = {978-3-642-24205-2},
}

@inproceedings{PoulsenNTV16,
  title = {Scopes Describe Frames: A Uniform Model for Memory Layout in Dynamic Semantics},
  author = {Casper Bach Poulsen and Pierre Néron and Andrew P. Tolmach and Eelco Visser},
  year = {2016},
  doi = {10.4230/LIPIcs.ECOOP.2016.20},
  url = {http://dx.doi.org/10.4230/LIPIcs.ECOOP.2016.20},
  researchr = {https://researchr.org/publication/PoulsenNTV16},
  cites = {0},
  citedby = {0},
  booktitle = {30th European Conference on Object-Oriented Programming, ECOOP 2016, July 18-22, 2016, Rome, Italy},
  editor = {Shriram Krishnamurthi and Benjamin S. Lerner},
  volume = {56},
  series = {LIPIcs},
  publisher = {Schloss Dagstuhl - Leibniz-Zentrum fuer Informatik},
  isbn = {978-3-95977-014-9},
}

@article{ThibaultCLMM00,
  title = {Static and Dynamic Program Compilation by Interpreter Specialization},
  author = {Scott Thibault and Charles Consel and Julia L. Lawall and Renaud Marlet and Gilles Muller},
  year = {2000},
  tags = {interpreter},
  researchr = {https://researchr.org/publication/ThibaultCLMM00},
  cites = {0},
  citedby = {0},
  journal = {Higher-Order and Symbolic Computation},
  volume = {13},
  number = {3},
  pages = {161-178},
}

@article{AntwerpenPRV18,
  title = {Scopes as types},
  author = {Hendrik van Antwerpen and Casper Bach Poulsen and Arjen Rouvoet and Eelco Visser},
  year = {2018},
  doi = {10.1145/3276484},
  url = {https://doi.org/10.1145/3276484},
  researchr = {https://researchr.org/publication/AntwerpenPRV18},
  cites = {0},
  citedby = {0},
  journal = {Proceedings of the ACM on Programming Languages},
  volume = {2},
  number = {OOPSLA},
}

@article{PelsmaekerAPV22,
  title = {Language-parametric static semantic code completion},
  author = {Daniël A. A. Pelsmaeker and Hendrik van Antwerpen and Casper Bach Poulsen and Eelco Visser},
  year = {2022},
  doi = {10.1145/3527329},
  url = {https://doi.org/10.1145/3527329},
  researchr = {https://researchr.org/publication/PelsmaekerAPV22},
  cites = {0},
  citedby = {0},
  journal = {Proceedings of the ACM on Programming Languages},
  volume = {6},
  number = {OOPSLA},
  pages = {1-30},
}

@inproceedings{VerguTV19,
  title = {Scopes and Frames Improve Meta-Interpreter Specialization},
  author = {Vlad A. Vergu and Andrew P. Tolmach and Eelco Visser},
  year = {2019},
  doi = {10.4230/LIPIcs.ECOOP.2019.4},
  url = {https://doi.org/10.4230/LIPIcs.ECOOP.2019.4},
  researchr = {https://researchr.org/publication/VerguTV19},
  cites = {0},
  citedby = {0},
  booktitle = {33rd European Conference on Object-Oriented Programming, ECOOP 2019, July 15-19, 2019, London, United Kingdom},
  editor = {Alastair F. Donaldson},
  volume = {134},
  series = {LIPIcs},
  publisher = {Schloss Dagstuhl - Leibniz-Zentrum fuer Informatik},
  isbn = {978-3-95977-111-5},
}

@inproceedings{Futamura82,
  title = {Partial Computation of Programs},
  author = {Yoshihiko Futamura},
  year = {1982},
  tags = {partial evaluation, Futamura},
  researchr = {https://researchr.org/publication/Futamura82},
  cites = {0},
  citedby = {0},
  pages = {1-35},
  booktitle = {RIMS Symposium on Software Science and Engineering, Kyoto, Japan, 1982, Proceedings},
  editor = {Eiichi Goto and Koichi Furukawa and Reiji Nakajima and Ikuo Nakata and Akinori Yonezawa},
  volume = {147},
  series = {Lecture Notes in Computer Science},
  publisher = {Springer},
  isbn = {3-540-11980-9},
}

@article{HedinM03,
  title = {JastAdd--an aspect-oriented compiler construction system},
  author = {Görel Hedin and Eva Magnusson},
  year = {2003},
  doi = {10.1016/S0167-6423(02)00109-0},
  url = {http://dx.doi.org/10.1016/S0167-6423(02)00109-0},
  tags = {compiler, JastAdd},
  researchr = {https://researchr.org/publication/HedinM03},
  cites = {0},
  citedby = {2},
  journal = {Science of Computer Programming},
  volume = {47},
  number = {1},
  pages = {37-58},
}

@inproceedings{AntwerpenNTVW16,
  title = {A constraint language for static semantic analysis based on scope graphs},
  author = {Hendrik van Antwerpen and Pierre Néron and Andrew P. Tolmach and Eelco Visser and Guido Wachsmuth},
  year = {2016},
  doi = {10.1145/2847538.2847543},
  url = {http://doi.acm.org/10.1145/2847538.2847543},
  researchr = {https://researchr.org/publication/AntwerpenNTVW16},
  cites = {0},
  citedby = {0},
  pages = {49-60},
  booktitle = {Proceedings of the 2016 ACM SIGPLAN Workshop on Partial Evaluation and Program Manipulation, PEPM 2016, St. Petersburg, FL, USA, January 20 - 22, 2016},
  editor = {Martin Erwig and Tiark Rompf},
  publisher = {ACM},
  isbn = {978-1-4503-4097-7},
}

@inproceedings{BradyH10,
  title = {Scrapping your inefficient engine: using partial evaluation to improve domain-specific language implementation},
  author = {Edwin Brady and Kevin Hammond},
  year = {2010},
  doi = {10.1145/1863543.1863587},
  url = {http://doi.acm.org/10.1145/1863543.1863587},
  tags = {partial evaluation, domain-specific language},
  researchr = {https://researchr.org/publication/BradyH10},
  cites = {0},
  citedby = {0},
  pages = {297-308},
  booktitle = {Proceeding of the 15th ACM SIGPLAN international conference on Functional programming, ICFP 2010, Baltimore, Maryland, USA, September 27-29, 2010},
  editor = {Paul Hudak and Stephanie Weirich},
  publisher = {ACM},
  isbn = {978-1-60558-794-3},
}

@article{OwensRT09,
  title = {Regular-expression derivatives re-examined},
  author = {Scott Owens and John H. Reppy and Aaron Turon},
  year = {2009},
  doi = {10.1017/S0956796808007090},
  url = {http://dx.doi.org/10.1017/S0956796808007090},
  researchr = {https://researchr.org/publication/OwensRT09},
  cites = {0},
  citedby = {0},
  journal = {Journal of Functional Programming},
  volume = {19},
  number = {2},
  pages = {173-190},
}

@inproceedings{SzaboKBME18,
  title = {Incremental overload resolution in object-oriented programming languages},
  author = {Tamás Szabó and  Edlira Kuci and Matthijs Bijman and Mira Mezini and Sebastian Erdweg},
  year = {2018},
  doi = {10.1145/3236454.3236485},
  url = {https://doi.org/10.1145/3236454.3236485},
  researchr = {https://researchr.org/publication/SzaboKBME18},
  cites = {0},
  citedby = {0},
  pages = {27-33},
  booktitle = {Companion Proceedings for the ISSTA/ECOOP 2018 Workshops, ISSTA 2018, Amsterdam, Netherlands, July 16-21, 2018},
  editor = {Julian Dolby and William G. J. Halfond and Ashish Mishra},
  publisher = {ACM},
  isbn = {978-1-4503-5939-9},
}

@article{Brzozowski64,
  title = {Derivatives of Regular Expressions},
  author = {Janusz A. Brzozowski},
  year = {1964},
  researchr = {https://researchr.org/publication/Brzozowski64},
  cites = {0},
  citedby = {0},
  journal = {Journal of the ACM},
  volume = {11},
  number = {4},
  pages = {481-494},
}

@article{RouvoetAPKV20,
  title = {Knowing when to ask: sound scheduling of name resolution in type checkers derived from declarative specifications},
  author = {Arjen Rouvoet and Hendrik van Antwerpen and Casper Bach Poulsen and Robbert Krebbers and Eelco Visser},
  year = {2020},
  doi = {10.1145/3428248},
  url = {https://doi.org/10.1145/3428248},
  researchr = {https://researchr.org/publication/RouvoetAPKV20},
  cites = {0},
  citedby = {0},
  journal = {Proceedings of the ACM on Programming Languages},
  volume = {4},
  number = {OOPSLA},
}

@mastersthesis{Misteli21,
  title = {Renaming for Everyone: Language-parametric Renaming in Spoofax},
  author = {Misteli, Phil},
  year = {2021},
  month = may,
  url = {http://resolver.tudelft.nl/uuid:60f5710d-445d-4583-957c-79d6afa45be5},
  note = {Available at \url{http://resolver.tudelft.nl/uuid:60f5710d-445d-4583-957c-79d6afa45be5}.},
  cites = {0},
  citedby = {0},
  school = {Delft University of Technology}
}

@mastersthesis{Gugten22,
  title = {Function Inlining as a Language Parametric Refactoring},
  author = {Van der Gugten, Loek},
  year = {2022},
  month = jun,
  url = {http://resolver.tudelft.nl/uuid:15057a42-f049-4321-b9ee-f62e7f1fda9f},
  note = {Available at \url{http://resolver.tudelft.nl/uuid:15057a42-f049-4321-b9ee-f62e7f1fda9f}.},
  cites = {0},
  citedby = {0},
  school = {Delft University of Technology}
}

@misc{JMH,
  author = {{OpenJDK}},
  title = {{Java} {Microbenchmark} {Harness} ({JMH})},
  url = {https://openjdk.java.net/projects/code-tools/jmh/},
  year = {2021}
}

@misc{Graal,
  author = {Oracle},
  title = {Graal project},
  url = {https://www.graalvm.org/},
  year = {2021}
}

\iftoggle{extended}{
  \appendix
  \newpage

  \section{Example 1: Full Derivation}

This section contains a full derivation for Example 2 in \cref{subsec:query-resolution-example}.
The scope graph as depicted in \cref{fig:example-modules} is denoted as follows.
The set of scopes $S$ is:
\[
  S_{\SG} = \setOf{s_l, s_{l'}, s_x, s_f, s_{\lambda}, s_y }
\]
Similarly, the set of edges in this scope graph is
\begin{align*}
  E_{\SG} = \setOf{
      \edge{s_{l'}}{\lblLEX}{s_l},
      \edge{s_{\lambda}}{\lblLEX}{s_l}, \\
    & \edge{s_l}{\lblVAR}{s_x},
      \edge{s_{l'}}{\lblVAR}{s_f},
      \edge{s_{\lambda}}{\lblVAR}{s_y},
  }
\end{align*}
The data mapping is given by:
\begin{align*}
  \rho_{\SG}:{} & s_x \mapsto \id{x} : \tyNAT,
                  s_f \mapsto \id{f} : \tyFUN{\tyNAT}{\tyNAT}, 
                  s_y \mapsto \id{y} : \tyNAT
\end{align*}
Finally, the parameters to the query are:
\begin{align*}
             R &= \reclos{\lblLEX}\lblVAR \\
        \wfd_x &= \lambda d. \exists T. d \overset{?}{=} \mathsf{x} : T \\
       \lblOrd &= \setOf{ \lblVAR \lblLE \lblLEX } \\
       \equivd &= \lambda d_1 d_2. \top
\end{align*}
Here, the $\mathbf{D}$ is a predicate that holds if its argument $d$ has name $\mathsf{x}$ and an arbitrary type $T$.
The data equivalence condition $\equivd$ is trivially true ($\top$), to ensure all declaration can potentially shadow each other.

\subsection{Interpretation}

Now, we show how the query is computed by the query resolution algorithm.

\newcommand{\lblAll}{\setOf{ \lblLEX, \lblVAR }}
\newcommand{\epsLang}{\setOf{ \EOP }}
\newcommand{\resultPath}{\edge{\edge{s_{\lambda}}{\lblLEX}{s_l}}{\lblVAR}{s_x}}

\begin{figure*}[p]
\begin{minipage}{\textwidth}
\removelatexerror
\begin{algorithm}[H]
                    \Block{\ResolveAll{$\edge{s_{\lambda}}{\lblVAR}{s_y}$, $\reeps$}}{
                      $\rehead{\reeps} \cup \setOf{ \EOP \alt \reeps \in \langOf{\reeps}} = \epsLang$\;
                      \Block{\ResolveL{$\edge{s_{\lambda}}{\lblVAR}{s_y}$, $\epsLang$, $\reeps$}}{
                        $\maxOf{\epsLang} = \EOP$\;
                        $\smallerOf{\epsLang}{\EOP} = \emptyset$\;
                        \Block{\ResolveLl{$\edge{s_{\lambda}}{\lblVAR}{s_y}$, $\emptyset$, $\EOP$, $\reeps$}}{
                          \Block{\ResolveL{$\edge{s_{\lambda}}{\lblVAR}{s_y}$, $\emptyset$, $\reeps$}}{
                            \Return $\emptyset$
                          }
                          \Block{\ResolveEOP{$\edge{s_{\lambda}}{\lblVAR}{s_y}$}}{
                            $\wfd_x(\rho_{\SG}(s_y)) = \wfd_x(\id{y}) = \bot$\;
                            \Return $\emptyset$
                          }
                          \Block{\Shadow{$\emptyset$, $\emptyset$}}{
                            \Return $\emptyset$\;
                          }
                          \Return $\emptyset$\;
                        }
                      \Return $\emptyset$\;
                      }
                    \Return $\emptyset$\;
                    }
\end{algorithm}
\end{minipage}
\caption{Resolving Residual Query in $s_y$}
\label{fig:resq-sy}
\end{figure*}

\begin{figure*}[p]
\begin{minipage}{\textwidth}
\removelatexerror
\begin{algorithm}[H]
              \Block{\ResolveAll{$\edge{s_{\lambda}}{\lblLEX}{s_l}$, $\reclos{\lblLEX}\lblVAR$}}{
                $\rehead{\reclos{\lblLEX}\lblVAR} \cup \setOf{ \EOP \alt \reeps \in \langOf{\reclos{\lblLEX}\lblVAR}} = \lblAll$\;
                \Block{\ResolveL{$\edge{s_{\lambda}}{\lblLEX}{s_l}$, $\lblAll$, $\reclos{\lblLEX}\lblVAR$}}{
                  $\maxOf{\lblAll} = \setOf{\lblLEX}$\;
                  \Block{$\smallerOf{\lblAll}{\lblLEX} = \setOf{\lblVAR}$}{
                    \Block{\ResolveLl{$\edge{s_{\lambda}}{\lblLEX}{s_l}$, $\setOf{\lblVAR}$, $\lblLEX$, $\reclos{\lblLEX}\lblVAR$}}{
                      \Block{\ResolveL{$\edge{s_{\lambda}}{\lblLEX}{s_l}$, $\setOf{\lblVAR}$, $\reclos{\lblLEX}\lblVAR$}}{
                        $\maxOf{\setOf{\lblVAR}} = \setOf{\lblVAR}$\;
                        \Block{$\smallerOf{\setOf{\lblVAR}}{\lblVAR} = \emptyset$}{
                          \Block{\ResolveLl{$\edge{s_{\lambda}}{\lblLEX}{s_l}$, $\emptyset$, $\lblVAR$, $\reclos{\lblLEX}\lblVAR$}}{
                            \Block{\ResolveL{$\edge{s_{\lambda}}{\lblLEX}{s_l}$, $\emptyset$, $\reclos{\lblLEX}\lblVAR$}}{
                              \Return $\emptyset$\;
                            }
                            \Block{\Resolvel{$\edge{s_{\lambda}}{\lblLEX}{s_l}$, $\lblVAR$, $\reclos{\lblLEX}\lblVAR$}}{
                              $\derive{\lblVAR}{\reclos{\lblLEX}\lblVAR} = \reeps$\;
                              \Block{\ResolveAll{$\resultPath$, $\reeps$}}{
                                $\rehead{\reeps} \cup \setOf{ \EOP \alt \reeps \in \langOf{\reeps}} = \epsLang$\;
                                \Block{\ResolveL{$\resultPath$, $\epsLang$, $\reeps$}}{
                                  $\maxOf{\epsLang} = \EOP$\;
                                  \Block{$\smallerOf{\epsLang}{\EOP} = \emptyset$}{
                                  \Block{\ResolveLl{$\resultPath$, $\emptyset$, $\EOP$, $\reeps$}}{
                                    \Block{\ResolveL{$\resultPath$, $\emptyset$, $\reeps$}}{
                                      \Return $\emptyset$
                                    }
                                    \Block{\ResolveEOP{$\resultPath$}}{
                                      $\wfd_x(\rho_{\SG}(s_x)) = \wfd_x(\id{x}) = \top$\;
                                      \Return $\setOf{\resultPath}$
                                    }
                                    \Block{\Shadow{$\emptyset$, $\setOf{\resultPath}$}}{
                                      \Return $\setOf{\resultPath}$\;
                                    }
                                    \Return $\setOf{\resultPath}$\;
                                  }
                                  }
                                  \Return $\setOf{\resultPath}$\;
                                }
                                \Return $\setOf{\resultPath}$\;
                              }
                              \Return $\setOf{\resultPath}$\;
                            }
                            \Block{\Shadow{$\emptyset$, $\setOf{\resultPath}$}}{
                              \Return $\setOf{\resultPath}$\;
                            }
                            \Return $\setOf{\resultPath}$\;
                          }
                          \Return $\setOf{\resultPath}$\;
                        }
                      }
                      \Return $\setOf{\resultPath}$\;
                    }
                    \Block{\Resolvel{$\edge{s_{\lambda}}{\lblLEX}{s_l}$, $\lblLEX$, $\reclos{\lblLEX}\lblVAR$}}{
                      \Return $\emptyset$
                    }
                    \Block{\Shadow{$\setOf{\resultPath}$, $\emptyset$}}{
                      \Return $\setOf{\resultPath}$\;
                    }
                  }
                  \Return $\setOf{\resultPath}$\;                
                }
                \Return $\setOf{\resultPath}$\;
              }
\end{algorithm}
\end{minipage}
\caption{Resolving Residual Query in $s_l$}
\label{fig:resq-sl}
\end{figure*}

\begin{figure*}[p]
\begin{minipage}{\textwidth}
\removelatexerror
\begin{algorithm}[H]
  \Block{\Resolve{$\SG$, $s_{\lambda}$, $\reclos{\lblLEX}\lblVAR$, $\wfd$, $\lblOrd$, $\equivd$}}{
    \Block{\ResolveAll{$s_{\lambda}$, $\reclos{\lblLEX}\lblVAR$}}{
      $\rehead{\reclos{\lblLEX}\lblVAR} \cup \setOf{ \EOP \alt \reeps \in \langOf{\reclos{\lblLEX}\lblVAR}} = \lblAll$\;
      \Block{\ResolveL{$s_{\lambda}$, $\lblAll$, $\reclos{\lblLEX}\lblVAR$}}{
        $\maxOf{\lblAll} = \setOf{\lblLEX}$\;
        \Block{$\smallerOf{\lblAll}{\lblLEX} = \setOf{\lblVAR}$}{
          \Block{\ResolveLl{$s_{\lambda}$, $\setOf{\lblVAR}$, $\lblLEX$, $\reclos{\lblLEX}\lblVAR$}}{
            \Block{\ResolveL{$s_{\lambda}$, $\setOf{\lblVAR}$, $\reclos{\lblLEX}\lblVAR$}}{
              $\maxOf{\setOf{\lblVAR}} = \setOf{\lblVAR}$\;
              \Block{$\smallerOf{\setOf{\lblVAR}}{\lblVAR} = \emptyset$}{
                \Block{\ResolveLl{$s_{\lambda}$, $\emptyset$, $\lblVAR$, $\reclos{\lblLEX}\lblVAR$}}{
                  \Block{\ResolveL{$s_{\lambda}$, $\emptyset$, $\reclos{\lblLEX}\lblVAR$}}{
                    \Return $\emptyset$\;
                  }
                  \Block{\Resolvel{$s_{\lambda}$, $\lblVAR$, $\reclos{\lblLEX}\lblVAR$}}{
                    $\derive{\lblVAR}{\reclos{\lblLEX}\lblVAR} = \reeps$\;
                    \Block{\ResolveAll{$\edge{s_{\lambda}}{\lblVAR}{s_y}$, $\reeps$}}{
                      \Return $\emptyset$ \tcp{See \cref{fig:resq-sy}}
                    }
                    \Return $\emptyset$\;
                  }
                  \Block{\Shadow{$\emptyset$, $\emptyset$}}{
                    \Return $\emptyset$\;
                  }
                  \Return $\emptyset$\;
                }
              }
              \Return $\emptyset$\;
            }
            \Block{\Resolvel{$s_{\lambda}$, $\lblLEX$, $\reclos{\lblLEX}\lblVAR$}}{
              $\derive{\lblLEX}{\reclos{\lblLEX}\lblVAR} = \reclos{\lblLEX}\lblVAR$\;
              \Block{\ResolveAll{$\edge{s_{\lambda}}{\lblLEX}{s_l}$, $\reclos{\lblLEX}\lblVAR$}}{
                \Return $\setOf{\resultPath}$ \tcp{see \cref{fig:resq-sy}}
              }
            }
            \Block{\Shadow{$\emptyset$}{$\setOf{\resultPath}$}}{
              \Return $\setOf{\resultPath}$
            }
            \Return $\setOf{\resultPath}$
          }
        }
        \Return $\setOf{\resultPath}$
      }
      \Return $\setOf{\resultPath}$
    }
    \Return $\setOf{\resultPath}$
  }
\end{algorithm}
\end{minipage}
\caption{Resolution Trace of Query Algorithm for Example 1.}
\label{fig:query-slam}
\end{figure*}
}{
}

\end{document}